\documentclass[12pt,a4paper,twoside]{article}
\usepackage{times}
\usepackage[utf8]{inputenc}
\usepackage{url}
\usepackage{scalefnt}
\usepackage{fancyhdr}
\usepackage{amsthm}
\usepackage{amsmath}
\usepackage{amssymb}
\usepackage{amsfonts,amssymb,amsxtra}
\usepackage[mathcal]{euscript}
\usepackage{psfrag}
\usepackage{graphicx}
\usepackage{lipsum}
\usepackage{bbm}
\usepackage{color, colortbl}
\usepackage{tabularx}
\usepackage{longtable} 
\usepackage{epsfig}
\usepackage{float}
\usepackage{tikz}
\usetikzlibrary{shapes.geometric, arrows}
\usepackage{import}
\usepackage{mathptmx}
\usepackage[scaled=0.92]{helvet}
\usepackage{lastpage}
\usepackage{setspace}
\usepackage{geometry}
\usepackage{indentfirst}
\usepackage{color}
\usepackage{colortbl}
\usepackage{subcaption}
\usepackage{multicol}
\usepackage{multirow}
\usepackage{arydshln}
\usepackage{natbib}
\usepackage[titletoc]{appendix}
\usepackage{booktabs}
\usepackage{caption}
\usepackage{dsfont}
\usepackage{listings}
\usepackage{mathtools}
\usepackage{amsmath}
\usepackage{hyperref}
\usepackage{bbm}
\usepackage[figuresright]{rotating}
\usepackage{makecell}
\usepackage{nccmath}

\tikzstyle{arrow} = [thick,->,>=stealth]
\tikzstyle{startstop} = [rectangle, minimum width=8cm, minimum height=1cm,text centered, draw=black]
\definecolor{gray}{rgb}{0.9,0.9,0.9}
\definecolor{coral}{rgb}{1.0, 0.5, 0.31}
\definecolor{coral2}{rgb}{0.93,0.42,0.31}
\definecolor{cyan3}{rgb}{0,0.80,0.80}
\definecolor{darkred}{rgb}{0.55,0,0}
\definecolor{green4}{rgb}{0,0.55,0}
\definecolor{tan4}{rgb}{0.55,0.35,0.17}
\definecolor{slategrey}{rgb}{0.44,0.5,0.56}
\definecolor{cadetblue4}{rgb}{0.33,0.53,0.55}

\definecolor{indianred3}{RGB}{205,85,85}
\definecolor{darkslategray4}{RGB}{82,139,139}

\begin{document}

\begin{center}
\large{\textbf{\textbf{A Zero-Inflated Spatio-Temporal Model for Integrating Fishery-Dependent and Independent Data under Preferential Sampling}}}
\end{center}
\vspace{0.2cm}

\begin{center}
Daniela Silva*\\
Division of Modeling and Management of Fishery Resources, Portuguese Institute for the Sea and Atmosphere (IPMA), Lisbon, Portugal,\\
Centre of Mathematics, University of Minho, Braga, Portugal,\\
\textit{daniela.dasilva@ipma.pt}\\
\vspace{0.3cm}

Raquel Menezes\\
Centre of Mathematics, University of Minho, Guimarães, Portugal
\vspace{0.3cm}

Gonçalo Araújo\\
Nova School of Business and Economics, Nova University Lisbon, Lisbon, Portugal,\\
Centre of Marine Sciences (CCMar), University of Algarve, Faro, Portugal,\\
University of Algarve, Faro, Portugal
\vspace{0.3cm}

Ana Machado\\
Instituto Dom Luiz (IDL), Faculty of Sciences, University of Lisbon, Lisbon, Portugal
\vspace{0.3cm}

Renato Rosa\\
Centre of Business and Economics Research, University of Coimbra, Coimbra, Portugal
\vspace{0.3cm}

Ana Moreno\\
Division of Modeling and Management of Fishery Resources, Portuguese Institute for the Sea and Atmosphere (IPMA), Lisbon, Portugal
\vspace{0.3cm}

Alexandra Silva\\
Division of Modeling and Management of Fishery Resources, Portuguese Institute for the Sea and Atmosphere (IPMA), Lisbon, Portugal
\vspace{0.3cm}

Susana Garrido\\
Division of Modeling and Management of Fishery Resources, Portuguese Institute for the Sea and Atmosphere (IPMA), Lisbon, Portugal
\end{center}
\vspace{0.2cm}

\begin{center}
\section*{Abstract}
\end{center}

Sustainable management of marine ecosystems is vital for maintaining healthy fishery resources, and benefits from advanced scientific tools to accurately assess species distribution patterns. In fisheries science, two primary data sources are used: fishery-independent data (FID), collected through systematic surveys, and fishery-dependent data (FDD), obtained from commercial fishing activities. While these sources provide complementary information, their distinct sampling schemes - systematic for FID and preferential for FDD - pose significant integration challenges.

This study introduces a novel spatio-temporal model that integrates FID and FDD, addressing challenges associated with zero-inflation and preferential sampling (PS) common in ecological data. The model employs a six-layer structure to differentiate between presence-absence and biomass observations, offering a robust framework for ecological studies affected by PS biases. Simulation results demonstrate the model's accuracy in parameter estimation across diverse PS scenarios and its ability to detect preferential signals.

Application to the study of the distribution patterns of the European sardine populations along the southern Portuguese continental shelf illustrates the model's effectiveness in integrating diverse data sources and incorporating environmental and vessel-specific covariates. The model reveals spatio-temporal variability in sardine presence and biomass, providing actionable insights for fisheries management. Beyond ecology, this framework offers broad applicability to data integration challenges in other disciplines.\\

\textbf{Keywords:} Data integration; Fishery data; Preferential sampling;Species distribution model; Zero-inflated spatio-temporal modeling.\\

\section{Introduction}\label{sec:introd}

In recent decades, growing concern about the health of marine ecosystems has underscored the urgent need for social and political actions aimed at promoting sustainability. As a result, understanding biodiversity patterns and species dynamics has become an increasingly pressing concern in marine science. Developing robust scientific tools capable of identifying species distribution patterns and capturing fluctuations in their dynamics is critical to informing sustainable fisheries management and conservation practices, particularly in the context of climate change, leading species to shift their distributions beyond their historical ranges \citep{OLeary2022}.

Simultaneously, advances in technology and data collection methodologies have greatly expanded the availability and quality of ecological data. However, these developments also introduce significant analytical challenges, particularly in accounting for spatial and temporal variability \citep{Minaya2018} since data is collected using different methods which are typically gathered across different spatial locations and/or time points.

Within fisheries science, two main types of data are collected: fishery-independent data (FID) and fishery-dependent data (FDD). Both sources are crucial for managing fisheries, setting catch limits, and developing conservation strategies. FID refers to information collected independently of fishing activities, often through surveys or research programs specifically designed to assess fish populations. These surveys typically employ standardized sampling techniques, providing a more reliable assessment of fish population size, abundance and distribution \citep{Ault1998}, as it is not influenced by fishing behavior or economic interests. In contrast, FDD is derived from data collected during commercial fishing activities, such as logbooks, fishery surveys, or monitoring programs. This type of data provides valuable insights into the characteristics of fish catches, including size distribution, species composition, and catch rates, and helps estimate fishing mortality and monitor fishing effort over time \citep{Rosenberg2005}.

Although FID generally offers more reliable estimates of fish population size, abundance, and distribution - being less affected by the biases inherent in commercial fishing - it is often constrained by limited temporal coverage due to high operational costs. In contrast, FDD is collected more frequently, due to the continuous nature of commercial activities, but is subject to spatial preferentiality, as fishermen typically concentrate their efforts in specific regions.

Because FID and FDD capture different facets of the underlying fish population dynamics - one unbiased but sparse, the other dense but biased - their integration into a single statistical framework can yield more robust inference. The benefits of combining multiple datasets have been well-documented across various research fields \citep{Steele2008, Kirk2012, Ferreras2021}, particularly in species distribution modeling \citep{Doser2021, Tehrani2022}. However, joint modeling of FID and FDD introduces additional complexities, related to sampling designs and preferential sampling (PS) biases, which are poorly handled by classical statistical tools. PS not only affects the predictive surface of spatial models but also leads to biased parameter estimates, if not properly accounted for \citep{Diggle2010, Gelfand2012}.

Previous work in this area has laid important groundwork for integrating species data. For example, \cite{Gelfand2019} proposed that PS can enable a probabilistically coherent fusion of presence-absence and presence-only data. While the authors highlighted the significant role of data fusion and the influence of PS, they did not handled the challenges posed by the distinct sampling designs from each data type, as seen with FID and FDD. Morever, their proposal was limited to modeling binary outcomes.

More recent models have specifically tackled the joint modeling of FID and FDD. For example, \citet{Rufener2021} introduced a three-layer hierarchical model combining scientific survey data with commercial catch data, incorporating distinct observation models and catchability parameters. Similarly, \citet{Alglave2022} developed a four-layer model using a shared latent biomass field, modeled with spatial random effects, and inhomogeneous Poisson processes (IPPs) to describe the sampling mechanisms. The degree of PS is quantified by the scaling parameter between the process of interest and the sampling process. This approach highlights the growing recognition of PS as a fundamental issue in joint modeling and species distribution inference.

Modeling fish distribution often requires accounting for spatial autocorrelation due to unobserved environmental variables or ecological interactions such as competition, dispersal, and aggregation \citep{Guelat2018}. Moreover, species abundance is inherently dynamic, varying across both space and time \citep{Hefley2016}, making it essential to incorporate temporal scales into SDMs \citep{Paradinas2017, Minaya2018}.

Despite this progress, challenges remain - particularly in accounting for zero-inflation, a common feature of ecological data due to true absences or non-detection. To address this, zero-inflated (ZI) models \citep{Lambert1992} and their hierarchical variants have become widely used in species distribution modeling \citep{MacKenzie2006,Arroita2012}, offering improved model fit and interpretability when excess zeros are present.

In this study, we propose a novel six-layer joint spatio-temporal model to infer spatio-temporal fish distribution capable of:  (i) distinguish between presence-absence and biomass/ {abundance} data; (ii) account for different observation processes across FID and FDD; (iii) incorporate PS effects in a structured way; and (iv) handle zero-inflation and spatial autocorrelation jointly. 

We evaluate the model's performance and challenges through a simulation study, exploring different PS scenarios and varying sample sizes for both FID and FDD. Finally, we apply our model to real-world data on the distribution of European sardine (\textit{Sardina pilchardus}, Walbaum 1792) along the Portuguese continental coast, using annual scientific acoustic survey data and commercial data collected from 2013 to 2018.

This work is structured into four sections. Section 2 details the proposed joint model, including its components, assumptions, and inference methodology. Section 3 outlines the simulation study, including the design and results. Section 4 presents the case study on European sardine distribution and its corresponding results. Section 5 discusses the findings and implications and explores potential avenues for future research. 

\section{Methods}

\subsection{Proposed joint model}
\label{ssec:jmodel}

\begin{figure}[h]
    \centering
    \includegraphics[scale=0.4]{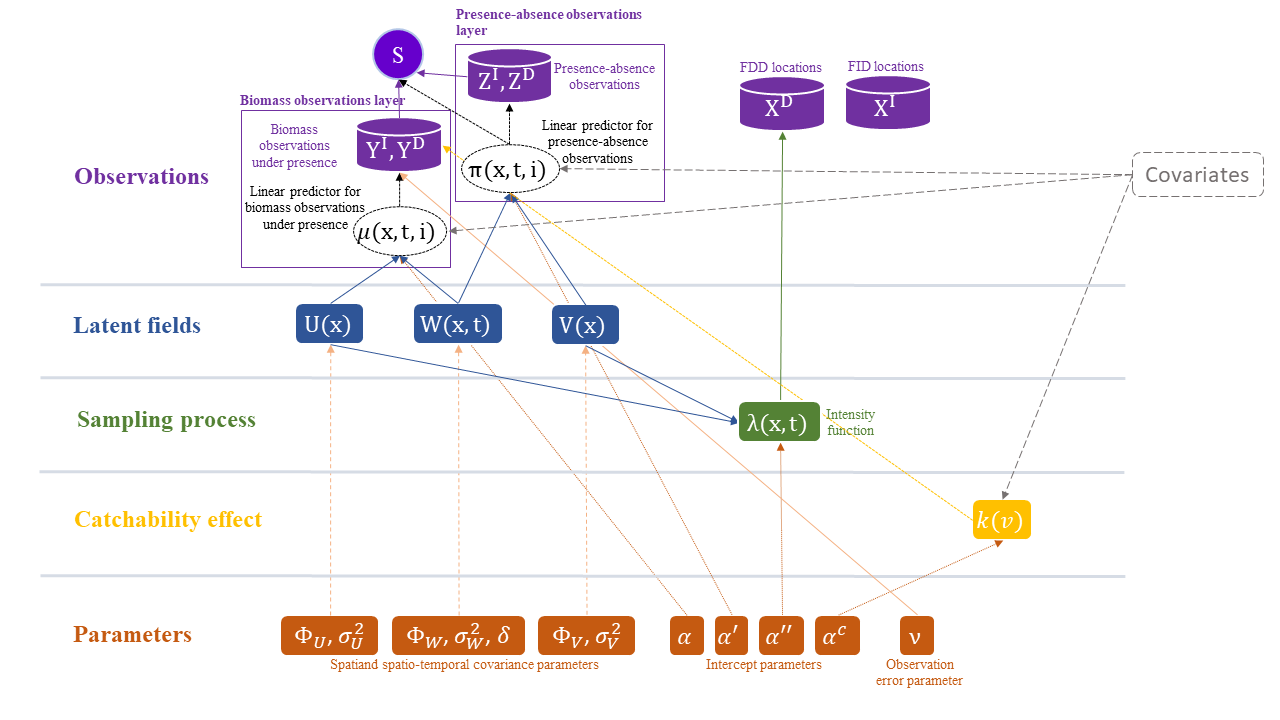}
    \caption{Diagram of the joint model, including the preferential sampling (PS) for fishery-dependent data (FDD), and distinguishing between presence–absence and biomass observations to address zero-inflated (ZI) data.}
    \label{fig:dag}
\end{figure}

To infer species distribution by leveraging the information from FID and FDD sources, we propose a spatio-temporal joint hierarchical model comprising six layers: presence-absence observations, biomass observations conditional on presence, sampling process, latent fields, catchability effects, and model parameters (Figure \ref{fig:dag}). This approach extends the spatial joint model introduced by \cite{Silva2025b} into a spatio-temporal framework that enables the temporal analysis of the phenomenon of interest. Given the potential distinct biomass indices derived from FID and FDD sources, as well as variations in vessel-specific catchability effects, the proposed model can be employed to evaluate the relative biomass index, which is proportional to the observed or captured biomass index (see Section \ref{sssection:catchability}).

\subsubsection{Observations}

Let us denote the spatio-temporal biomass process at time $t = \{t_1, \cdots, t_T\}$ and location $\mathbf{x}_i \in \mathcal{A} \subset \mathbb{R}^2$ by $\mathbf{S}=\{s(\mathbf{x}_1,t_1),\cdots,s(\mathbf{x}_{n(t_1)},t_1),\cdots,s(\mathbf{x}_{n(t_T)},t_T)\}$, where $\mathcal{A}$ is the study region and $n(t)$ represents the dimension of the data for time $t$. The PAP $\mathbf{Z}=\{z(\mathbf{x}_1,t_1),\cdots,z(\mathbf{x}_{n(t_1)},t_1),\cdots,$ $z(\mathbf{x}_{n(t_T)},t_T)\}$, with presence probability $\pi(\mathbf{x},t)$, takes the binary value 0 if no species was observed at location $\mathbf{x}$ and time $t$, and 1 otherwise. The biomass process under the presence $\mathbf{Y}=\mathbf{S} \vert (\mathbf{Z}=1)=\{y(\mathbf{x}_1,t_1),\cdots,y(\mathbf{x}_{n(t_1)},t_1),\cdots,y(\mathbf{x}_{n(t_T)},t_T)\}$ takes strictly positive values. 

The relative biomass process, denoted by $\mathbf{S}^*$, is proportional to the underlying biomass process $\mathbf{S}$, while the relative biomass process conditional on presence, denoted by $\mathbf{Y}^*$, is, analogously, proportional to $\mathbf{Y}$.

In accordance with \citet{Silva2024}, the distribution of the process of interest $\mathbf{S}$ is given by the product of the distribution of the PAP $\mathbf{Z}$ and the distribution of the biomass process under the presence $\mathbf{Y}$ such that
\begin{equation}
P(S(\mathbf{x},t)=s(\mathbf{x},t))= \left \{ \begin{matrix} 1-\pi(\mathbf{x},t)&~\text{if}~&s(\mathbf{x},t)=0\\ 
\pi(\mathbf{x},t)~p(s(\mathbf{x},t) \vert \zeta(\mathbf{x},t))&~\text{if}~&s(\mathbf{x},t)>0
\end{matrix}\right.
\label{eq_theory_hurdle}
\end{equation}
where $p(s(\mathbf{x},t) \vert \zeta(\mathbf{x},t))$ represents a probability mass function for the process of interest $S_{\mathbf{x}t}$, either truncated (e.g., Gamma and Log-normal distributions) or untruncated (e.g., Poisson and Negative Binomial distributions). The property of the product can be also applied to the main statistics of the process of interest, mean $E[\mathbf{S}]=E[\mathbf{Z}]E[\mathbf{Y}]$ and median $F_{\mathbf{S}}^{-1}(0.5)=E[\mathbf{Z}]F_{\mathbf{Y}}^{-1}(0.5)$ \citep{Silva2024}. An analogous decomposition holds for the distributional properties of the relative biomass process $\mathbf{S}^*$.

We propose a two-part model (\eqref{eq:logit} and \eqref{eq:log}) designed for the inference of species biomass distribution. This model is specifically crafted to accommodate ZI data, taking into account the distinct conditions influencing both the PAP \eqref{eq:logit} and the biomass process in the presence of the species \eqref{eq:log}. PAP $\mathbf{Z}$ is assumed to come from a Bernoulli distribution with probability $\pi$ such that $Z(\mathbf{x}t) \sim Bernoulli(\pi(\mathbf{x},t))$. The biomass process under the presence $\mathbf{Y}$ requires a continuous distribution such as Gamma distribution with shape parameter $a(\mathbf{x},t)=\zeta(\mathbf{x},t)^2/\upsilon^2$ and scaling parameter $b(\mathbf{x},t)=\upsilon^2/\zeta(\mathbf{x},t)$, that is, $Y(\mathbf{x},t) \sim Gamma(a(\mathbf{x},t),b(\mathbf{x},t))$. $\zeta(\mathbf{x},t)$ represents the mean biomass at location $\mathbf{x}$ and time $t$, while $\upsilon$ denotes the corresponding standard deviation, both conditional on species presence. Specifically, $\mu(\mathbf{x},t)$ defines the expected relative biomass process under presence, $Y(\mathbf{x},t)^*$, at location $\mathbf{x}$ and time $t$. The relationship between $\upsilon(\mathbf{x},t)$ and $\mu(\mathbf{x},t)$ is detailed in \eqref{eq:relat_biom}.
\begin{align}
logit(\pi(\mathbf{x},t,i))&=\alpha' + \sum_{j=1}^{p'}f'(K(C'(j,\mathbf{x},t,i),c,l)) + V(\mathbf{x}) + W(\mathbf{x},t) \label{eq:logit}\\
    log(\mu(\mathbf{x},t,i)) &= \alpha + \sum_{j=1}^{p}f(K(C(j,\mathbf{x},t,i),c,l)) + U(\mathbf{x}) + W(\mathbf{x},t) \label{eq:log}
\end{align}

$\alpha$ and $\alpha'$ parameters denote the intercepts of the linear predictors for the corresponding process. $f(.)$ and $f'(.)$ denote linear or smoother (for instance, B-splines, thin plate, and cubic regression splines) effects of covariates $C(.)$ and $C'(.)$, respectively. $K(.)$ represents a weighted average of the covariates \citep{Silva2024}, enabling the incorporation of short-term time-lagged effects of the covariates. The index $i$ refers to the $i$-th subperiod (e.g., day) within time $t$ (e.g, year), thereby facilitating the modeling of recent temporal influences. The spatial latent fields are denoted by $U(\mathbf{x})$ and $V(\mathbf{x})$, while $W(\mathbf{x},t)$ represents the spatio-temporal field, as described below.

\begin{figure}[H]
\centering
\includegraphics[scale=0.65]{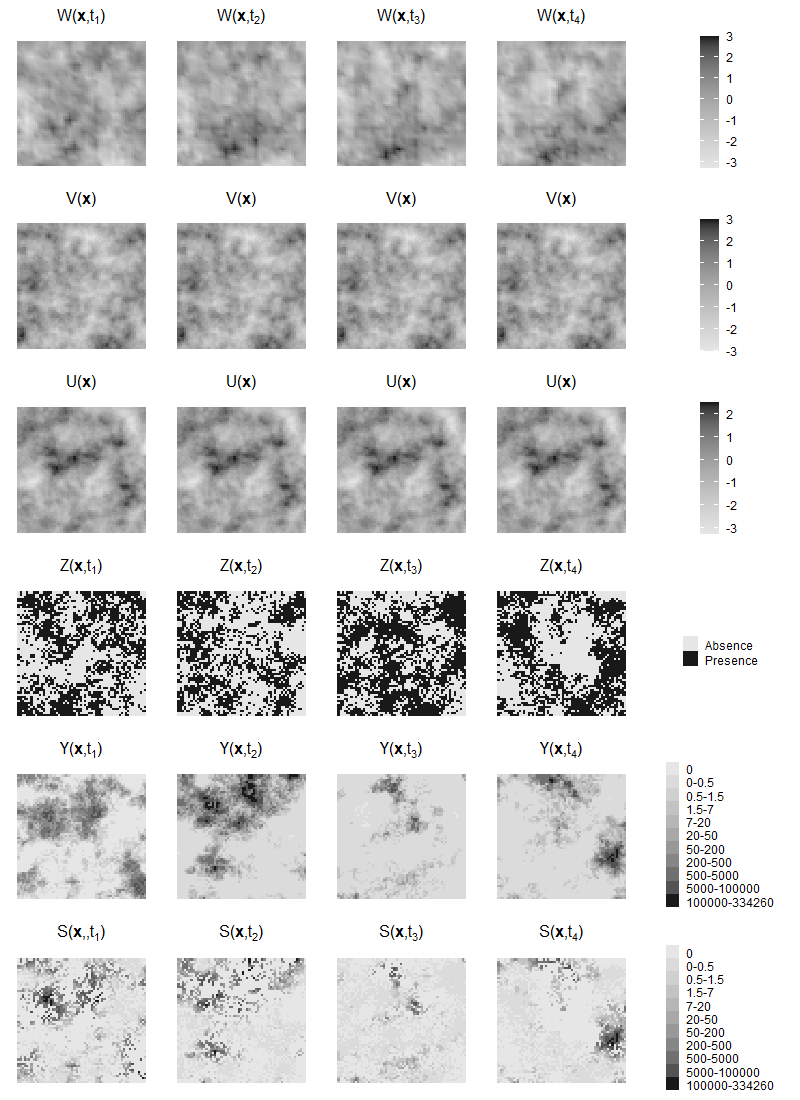}
\caption{Example of simulated latent fields. ${V}({\mathbf{X}})$ and ${U}(\mathbf{X})$ represent realizations of the simulated GMRFs, while $\mathbf{W}(\mathbf{x},t)$ are the realizations of simulated spatio-temporal field with temporal correlation $\delta=0.8$ and $t=1,\cdots, T$ where $T=4$. The biomass process ${S}(\mathbf{x},t)$ is derived as ${S}(\mathbf{x},t) = {Z}(\mathbf{x},t) \cdot {Y}(\mathbf{x},t)$, where ${Y}(\mathbf{x},t)$ represents realizations of the biomass process under presence determined by $\zeta(\mathbf{x},t)$. The ${Z}(\mathbf{x},t)$ represents realizations of the PAP determined by $\pi(\mathbf{x},t)$. The $\pi(\mathbf{x},t)$ and $\zeta(\mathbf{x},t)$ are the probability field of species presence and mean field of species biomass under presence given ${V}(\mathbf{X})$ and ${U}(\mathbf{X})$, respectively, and given ${W}(\mathbf{x},t)$.}
\label{fig:simu_data}
\end{figure}

Figure \ref{fig:simu_data} illustrates a realization of each process contributing to the determination of a realization of the process $\mathbf{S}$, as described above.

\subsubsection{Latent fields}
\label{sssec:latent}

Following \citet{Silva2025b}, the biomass is modeled using spatial fields $\mathbf{V}$ and $\mathbf{U}$, each corresponding to the processes $\mathbf{Z}$ and $\mathbf{Y}$, respectively. Additionally, a spatio-temporal field $\mathbf{W}$ is introduced to account for the dynamic structure shared between $\mathbf{Y}$ and $\mathbf{Z}$.

Underlying the spatio-temporal phenomenon there is a precision matrix $\mathbf{Q}$, governed by the structure matrix $\mathbf{R}_W$, where $\mathbf{Q} = \tau_W\mathbf{R}_W$ and $\tau_W$ is an unknown scalar. Consequently, the structure matrix delineates the nature of temporal and/or spatial interdependencies among elements of $\mathbf{W}$, and it can be factorized as the Kronecker product of structure matrices of the corresponding interacting main effects. In our analysis, the phenomenon exhibits temporal change following a first-order autoregressive process with temporal correlation $\delta$, as defined by the equation:
\begin{equation}
    W(\mathbf{x},t) = \delta W(\mathbf{x},t-1) + \xi(\mathbf{x},t)
\end{equation}
where $t = \{t_2, \cdots, t_T\}$, $ \vert \delta \vert < 1$, and $ W(\mathbf{x},t_1) \sim N(\mathbf{0},\sigma_W^2/(1-\delta^2))$. 

Meanwhile, $\xi(\mathbf{x},t)$ denotes a zero-mean Gaussian Field (GF), assumed to be temporally independent. Thus, $Cov(\xi(\mathbf{x}_i,t),\xi(\mathbf{x}_j,h))=Cov(\xi(\mathbf{x}_i),\xi(\mathbf{x}_j))$ for $t = h$ and $\mathbf{x}_i \neq \mathbf{x}_j$.

To allow both processes $\mathbf{Y}$ and $\mathbf{Z}$ to display distinct spatial structures, two spatial fields $\mathbf{U}(\mathbf{X})$ and ${V}(\mathbf{X})$ are added to each linear predictor \eqref{eq:logit} and \eqref{eq:log}. Each spatial latent field (${U}(\mathbf{X}),$ ${V}(\mathbf{X})$ and $\xi(\mathbf{X})$) denotes the spatial dependency and variation that is accounted for through a zero-mean Gaussian Markov Random Field (GMRF) with a Mat\'{e}rn covariance function $M(\mathbf{x},\mathbf{x'};\phi_.,\sigma_., \nu)$ with spatial range $\phi_.$, marginal variance $\sigma_.^2$ and smoothing parameter $\nu$ such that ${U}(\mathbf{X}),{V}(\mathbf{X}),\xi(\mathbf{X}) \sim GMRF(\mathbf{0},M(\mathbf{x},\mathbf{x'};\phi_l,\sigma_l,\nu)),~l=\{\mathbf{U},\mathbf{V},\mathbf{W}\}$ where we can consider $\phi_{\mathbf{W}}=\phi_{\mathbf{\xi}}$ and $\sigma_{\mathbf{W}}=\sigma_{\mathbf{\xi}}$.

\subsubsection{Sampling process}

In a spatio-temporal setting, let us denote the spatial point processes underlying FID and FDD by $\mathbf{X}^{I}(t)$ and $\mathbf{X}^{D}(t)$ for time $t$, respectively\footnote{$I$ is the acronym for FID in $\mathbf{X}^{I}(t)$ and $D$ for FDD in $\mathbf{X}^{D}(t)$.}. The intensity of a point process can exhibit either spatial constancy, yielding a homogeneous or stationary pattern, or spatial variability with a discernible spatial trend, resulting in an inhomogeneous pattern. However, the assumption of stationarity may prove unrealistic in certain applications. This is particularly evident when the process of interest dictates the data locations $\mathbf{x}^{D}$, and there exists stochastic dependence between this process and the one under consideration, as is often the case in FDD. 

FID typically enable a more objective assessment of species distribution, as the sampling locations $\mathbf{x}^I$ for time $t$ are selected independently of the latent ecological process $\mathbf{S}(t)$, commonly via randomized or systematic sampling designs. When the design is spatially uniform and randomized, $\mathbf{X}^I$ can be modeled as a realization of a Homogeneous Poisson Process (HPP) with constant intensity $\lambda^{HPP}$, such that 
$\mathbf{X}^{I}(t) \sim HPP(\lambda^{HPP}(t))$. In such cases, since the sampling mechanism is independent of the ecological process, its contribution to the joint likelihood is constant and does not affect parameter estimation. Alternatively, under systematic sampling designs, the locations $\mathbf{X}^I$ are fixed and treated as deterministic within the inferential framework. In both scenarios, the sampling design does not convey information about the latent ecological field; thus, $\mathbf{X}^I$ is either excluded from the likelihood function or conditioned upon during model fitting. This approach ensures valid statistical inference while accommodating diverse FID sampling schemes.

Following \cite{Diggle2010}, the set of fishing locations $\mathbf{X}^{D}(t)$ at time $t$ is modeled conditionally on ${U}(\mathbf{X})$ and ${V}(\mathbf{X})$ as an IPP with intensity function $\lambda(\mathbf{x}^{D},t)$, $\mathbf{X}^{D}(t) \sim IPP(\lambda(\mathbf{x}^{D},t))$. For clarity and notational simplicity, we hereafter assume $\mathbf{X}(t)=\mathbf{X}^{D}(t)$. Therefore, the logarithm of the intensity function is expressed as:
\begin{align}
    log(\lambda(\mathbf{x},t))=\alpha''(t) + &\beta'(t) V({\mathbf{x}}) + \beta(t) U({\mathbf{x}}).
    \label{eq:ipp}
\end{align}
The intensity function of the IPP for time $t$ is described by the logarithmic link function of the linear combination of the intercept $\alpha''_t$ and the latent effects $U(\mathbf{x}^{D})$ and $V(\mathbf{x}^{D})$.

For a given time $t$, the parameters $\beta'(t)$ and $\beta(t)$ quantify the degree of spatial PS by scaling the relationship between local fishing intensity and the local values of the processes of interest, ${Z}({.,t})$ and ${Y}({.,t})$, respectively. Building upon \citet{Silva2025b}, both preferential parameters $\beta(t)$ and $\beta'(t)$ are allowed to be dynamic and vary over time to illustrate the capture of the distinct sampling intensities inherent to each time $t$. These sampling intensities may fluctuate in time due to various factors, such as changes in fishery quotas and regulatory restrictions.

\subsubsection{Catchability effects}
\label{sssection:catchability}

To address potentially distinct biomass indices derived from different fishery data sources and variations in vessel-specific catchability effects, we define the expected biomass index for each vessel $v$ (whether associated with FID or FDD), spatial location $\mathbf{x}$, time $t$ and subperiod $i$, denoted as $\zeta(\mathbf{x},t,i,v)$. This index is modeled as the product of the expected relative biomass $\mu(\mathbf{x},t,i)$ and a vessel-specific catchability effect $k(v)$:
\begin{equation}
\zeta(\mathbf{x},t,i,v) = k(v) \times \mu(\mathbf{x},t,i).
\label{eq:relat_biom}
\end{equation}

The catchability associated with each commercial vessel varies due to multiple factors, including vessel characteristics (e.g., vessel length and engine power). To account for this variability, we model catchability $k(v)$ as a function of fixed effects $F$ (representing vessel attributes) and random effects $\gamma_c(v)$ (capturing vessel-specific variation), given by:
\begin{equation}
k(v) = \exp \left\{ \alpha_c + \sum_{h=1}^{H} f_c(F(h,v)) + \gamma_c(v) \right\}.
\label{eq:catch_effect}
\end{equation}

Here, $\alpha_c$ denotes the intercept, while $\gamma_c$ follows an independent and identically distributed (i.i.d.) Gaussian process with variance $\sigma_{\gamma_c}^2$. $f_c(.)$ represents linear or smoother effects of covarite $F(h,.)$. 

The inclusion of catchability effect, $k(v)$, enables adjustment for measurement discrepancies between the two data sources, accounting for vessel-specific differences in catch efficiency. This ensures that relative biomass estimates remain comparable across data sources, thereby maintaining proportionality to the underlying biomass.

\subsection{Inference and estimation}
\label{ssec:inf_est}

The model estimation and parameter inference are conducted through Laplace Approximation \citep{Skaug2006} performed with the Template Model Builder \citep{Kristensen2016} using the $\texttt{TMB~R}$ package. This method has superior performance in parameter estimation and latent field predictions compared to Monte Carlo approximation and non-preferential models, and offers greater flexibility in defining complex PS models compared to the INLA method \citep{Dinsdale2018}. Therefore, the $\texttt{C++}$ template function was built based on the likelihood of the proposed model presented in Section \ref{sssec:lik}, subsequently utilized for application through the $\texttt{TMB~R}$ package.

To derive each spatial latent field, we employ an approximation method based on stochastic partial differential equations (SPDEs), as introduced by \cite{Lindgren2011}. This SPDE approach enables the approximation of a spatial continuous field, represented by a Mat\'{e}rn covariance function, using a GMRF. The adoption of this approximation is motivated by its computational efficiency. Parameterization is performed regarding marginal variance $\sigma^2_.$ and range of influence $\phi_.$, enhancing the model interpretability and computational advantages.

Under specific circumstances, a reparametrization of the model parameters $\phi_.$, $\sigma_.$, and $\delta$ (as defined in Section \ref{sssec:latent}) proves to be advantageous. In this study, ditacted by the functionality of the $\texttt{TMB~R}$ package aiming the reduction of the optimization convergence time, our proposed model is implemented using the parameters $\kappa$, $\tau$ and $\delta^{*}$. Subsequently, assuming a fixed value of $\nu=1$, a reparametrization is undertaken to enhance the interpretability of the results according to $\phi_.=\frac{\sqrt{8\nu}}{\kappa_.}$, $\sigma_.=\frac{\sqrt{\Gamma(\nu)}}{\sqrt{\Gamma(\nu+1)} \times \kappa_.^{\nu} \times \tau_. \times \sqrt{4\pi}}$ for $l=\{\mathbf{U},\mathbf{V},\mathbf{W}\}$, and  $\delta=2 \times \frac{exp(\delta^{*})}{1+exp(\delta^{*})}-1$.

\subsubsection{Likelihood}
\label{sssec:lik}

The joint distribution \eqref{eq:dist} is determined by the distribution of the biomass process under presence (conditioned on the GMRF $\mathbf{U}$, the spatio-temporal field $\mathbf{W}$ and the sampling processes $\mathbf{X}_t^I$ and $\mathbf{X}_t$), the distribution of the presence-absence process (PAP, conditioned on the GMRF $\mathbf{V}$, the spatio-temporal field $\mathbf{W}$ and the sampling processes $\mathbf{X}_t^I$ and $\mathbf{X}_t$), the distributions of the sampling processes $\mathbf{X}_t$ (conditioned on the GMRFs $\mathbf{U}$ and $\mathbf{V}$) and $\mathbf{X}_t^I$, the distribution of the spatio-temporal field $\mathbf{W}$, and the distributions of both GMRFs $\mathbf{U}$ and $\mathbf{V}$. Each of these distributions is characterized by a specific expression, with the distribution of $\mathbf{X}_t^I$ remaining constant as it is a HPP.
\begin{align}
\left[\mathbf{Y},\mathbf{Z},\mathbf{X},\mathbf{X}^{S},\mathbf{U},\mathbf{V},\mathbf{W} \right]  = &
\left [\mathbf{Y} \vert \mathbf{X},\mathbf{X}^{S},\mathbf{U},\mathbf{W} \right ]  \left [\mathbf{Z} \vert \mathbf{X},\mathbf{X}^{S},\mathbf{V},\mathbf{W} \right ] \left [\mathbf{X},\mathbf{X}^{S},\mathbf{U},\mathbf{V},\mathbf{W} \right]\nonumber\\ = &\left [\mathbf{Y} \vert \mathbf{X},\mathbf{X}^{S},\mathbf{U},\mathbf{W} \right ]  \left [\mathbf{Z} \vert \mathbf{X},\mathbf{X}^{S},\mathbf{V},\mathbf{W} \right ] \left [ \mathbf{X} \vert \mathbf{U}, \mathbf{V} \right] \left [\mathbf{U}\right ]
\left [\mathbf{V}\right ] \left [\mathbf{W}\right ] \left [\mathbf{X}^{S}\right ] 
\label{eq:dist}
\end{align}

Given the result presented in \eqref{eq:dist} and denoting the space of parameters as $\Theta$, \eqref{eq:lik} provides the likelihood for the proposed model.
\begin{align}
\mathcal{L}(\Theta)=&\mathcal{L}(\zeta, \sigma; \mathbf{y}) \times \mathcal{L}(\pi; \mathbf{z}) \times \mathcal{L}(\lambda; \mathbf{x}) \times \mathcal{L}(\sigma_U, \phi_U) \times \mathcal{L}(\sigma_V, \phi_V) \times \mathcal{L}(\sigma_W, \phi_W, \delta)
\label{eq:lik}
\end{align}
where $\mathcal{L}(\zeta, \sigma; \mathbf{y})$ represents the likelihood for $\mathbf{Y} \vert \mathbf{X},\mathbf{X}^{S} ,\mathbf{U},\mathbf{W}$, the likelihood for $\mathbf{Z} \vert \mathbf{X},\mathbf{X}^{S},\mathbf{V},\mathbf{W}$ is denoted by $\mathcal{L}(\pi; \mathbf{z})$,  $\mathcal{L}(\lambda; \mathbf{x})$ identifies the likelihood for $\mathbf{X} \vert \mathbf{U}, \mathbf{V}$, the likelihood for $\mathbf{W}$ is denoted by $\mathcal{L}(\sigma_W, \phi_W, \delta)$, and  $\mathcal{L}(\sigma_U, \phi_U)$ and $\mathcal{L}(\sigma_V, \phi_V)$ represent the likelihoods for $\mathbf{U}$ and $\mathbf{V}$, respectively.

Hence, the joint log-likelihood $\ell(\Theta)$ of the model is obtained by summing the log-likelihood contributions of all component processes described in \eqref{eq:lik}.

The full expressions for each process's likelihoods and corresponding log-likelihoods are given in the Section 1 of the Supplementary Material \citep{Silva2025}.

\section{Simulation study}

A simulation study is carried out to evaluate the performance of the proposed model. Therefore, in order to realize if the model can illustrate different phenomena various data sources were firstly simulated considering distinct sampling scenarios, and how the model is sensitive to the data dimension each scenario was simulated under different data size.

\subsection{Scenarios of sampling}

Various combinations of the parameters $\beta'(t)$ and $\beta(t)$ give rise to distinct intensity functions of the point processes, and consequently to diverse sampling scenarios. These scenarios may range from extremes, where sampling is solely contingent on either interest process $\mathbf{Z}$ or $\mathbf{Y}$, to situations where it is dependent on both processes. This array of scenarios enables the projection of real-world situations in fishery science, as fishermen often concentrate their efforts on sampling based on their prior knowledge of the species. For instance, fishermen may seek areas with higher species abundance and simultaneously direct their efforts toward locations where the species is present. Another scenario arises when fishermen exclusively target regions where the species is present, without specific concern for the quantity captured. This may occur due to established fishing quotas and restrictions.

Below, we enumerate three scenarios whose representation of the sample locations is available in Figure 1 of the Supplementary Material \citep{Silva2025}.
\begin{itemize}
    \item \textit{Scenario 1:} Strong PS dependent on $\mathbf{Y}$\\
The sampling process for simulated FDD is entirely and strongly contingent on the biomass process under presence. Hence, $\beta'(t) = 0$ and $\beta(t) \sim N(2,0.25)$ with $t=\{t_1,\cdots,t_4\}$.
    \item \textit{Scenario 2:} Strong PS dependent on $\mathbf{Z}$\\
The sampling locations for simulated FDD are contingent on the PAP of the species. The preferentiality parameters are set such that $\beta'(t) \sim N(2,0.25)$ and $\beta(t) = 0$ with $t=\{t_1,\cdots,t_4\}$.
    \item \textit{Scenario 3:} Moderate PS dependent on $\mathbf{Z}$ and strong PS dependent on $\mathbf{Y}$\\
The sampling process for FDD is dependent on both processes of interest, with a higher weight assigned to the biomass process under presence $\mathbf{Y}$. In this setting, $\beta'(t) \sim N(1,0.25)$ and $\beta(t) \sim N(2,0.25)$ with $t=\{t_1,\cdots,t_4\}$.
\end{itemize}

\subsection{Simulation-estimation experiments}

Each scenario is simulated on a regular $60 \times 60$ grid within the domain $[0,1] \times [0,1]$ over a time series of dimension 4. Range and marginal variance parameters are individually set for each GMRF, to assess the model performance concerning distinct spatial dependencies of both responses $\mathbf{Z}$ and $\mathbf{Y}$. This assumes that different spatial processes govern both responses. Specifically, $(\phi_{\mathbf{V}},\sigma_{\mathbf{V}}^2)=(0.15,0.80)$, $(\phi_{\mathbf{U}},\sigma_{\mathbf{U}}^2)=(0.20,1.00)$ and $(\phi_{\mathbf{W}},\sigma_{\mathbf{W}}^2)=(0.20,1.00)$. The temporal correlation associated with the spatio-temporal structure $\mathbf{W}$ is defined as $\delta=0.8$. Additionally, all intercept parameters, namely $\alpha$, $\alpha'$, and $\alpha''(t)$, were assumed to be zero across scenarios.

This simulation experiment also incorporates the effects of simulated external covariates, with two covariates describing each process $\mathbf{Z}$ and $\mathbf{Y}$. The covariates are generated as GMRFs with the following specifications: $C_{1}' \sim GMRF(\mathbf{0}, M(\mathbf{x},\mathbf{x}',0.25,1.5,1))$, $C_{2}' \sim GMRF(\mathbf{0}, M(\mathbf{x},\mathbf{x}',$ $0.2,1,1))$, $C_{1} \sim GMRF(\mathbf{0}, M(\mathbf{x},\mathbf{x}',0.3,1.75,1))$, and $C_{2} \sim GMRF(\mathbf{0}, M(\mathbf{x},\mathbf{x}',0.15,2,1))$. The effects of these covariates on the responses are assumed to be linear, with the regression coefficients for $\mathbf{Z}$ given by $\mathbf{\theta}'= (\theta_1',\theta_2')=(-1,1.5)$ and for $\mathbf{Y}$ given by $\mathbf{\theta}= (\theta_1,\theta_2)=(3,-0.5)$.

To assess how the sample sizes of both FID ($n^{I}(t)$) and FDD ($n^{D}(t)$) influence the relative contribution of each data source, simulations are conducted with various combinations of sample sizes $Comb(n^{I}(t), n^{D}(t))$. These combinations are chosen to represent possible real-world situations, allowing the dimensions of both data sources to be equal or different.

The selected combinations include $Comb(100,100)$ for all time units $t=\{t_1,\cdots, t_4\}$, representing a scenario where both data sources share identical dimensions. Recognizing that, in practical situations, FDD often exhibits larger dimensions compared to FID due to factors such as financial constraints and time-intensive surveying, additional combinations are explored. These include $Comb(100,200)$ and a more asymmetric scenario $Comb(100,500)$ for all time units $t=\{t_1,\cdots, t_4\}$. Conversely, to account for common yet plausible scenarios where a greater emphasis on FID may arise due to fishery restrictions or limited interest in specific species by fishermen, a combination with larger FID dimensions is considered, denoted as $Comb(200,100)$ for all time units $t=\{t_1,\cdots, t_4\}$.

In summary, the selected combinations provide a detailed exploration of the interplay between sample sizes of FID and FDD, capturing realistic scenarios ranging from balanced dimensions to instances where one data source dominates due to practical constraints and ecological considerations. Furthermore, the simulation study does not incorporate the modeling of catchability effects, assuming the same biomass index for FID and FDD sources and that vessel-specific catchability differences are negligible. Consequently, modeling the relative biomass is not required in this context and $\mu(\mathbf{x},t)= \zeta(\mathbf{x},t)$ can be assumed in \eqref{eq:log}.

To ensure robustness, each scenario and configuration is repeated 100 times, allowing for the capture of variability among replicates. 

\subsection{Performance metrics}

The assessment of the estimation performance of the proposed model involves a comprehensive analysis of various model outputs. The evaluation encompasses all estimated parameters, including the intercept, preferential, fixed, and spatio-temporal covariance parameters, as well as the spatio-temporal predictions.

Given the potential for asymmetric distributions in spatial covariance parameters and time correlation across replicas, their estimation quality is performed through the identification of the mode and the 90\% confidence interval given the skew distributions of these parameters. Additionally, the performance of the preferential parameters is evaluated through the median and the 90\% confidence interval for the relative bias given by $\frac{\hat{\beta}-\beta}{\beta}$.

In addition to assessing parameters estimation, the predictive performance of the proposed method is thoroughly evaluated using three distinct metrics: RMSE, MAE, and the Hellinger distance \citep{Lecam1986}. These metrics provide a robust evaluation of the model's ability to generate spatio-temporal predictions that align closely with observed data. In this context, the Hellinger distance measures the similarity between the observed data and the predicted data distributions, ranging from 0, indicating equality, and 1, indicating ``total difference''.

\subsection{Results}

\subsubsection{Evaluation of the estimation of preferential parameters}

The proposed model demonstrates robust performance in estimating preferential parameters across various scenarios (Table \ref{tab:pref_results}). Specifically, when $\beta'(t)$ or $\beta(t)$ is assumed to be zero, as in \textit{Scenarios 1} and \textit{2}, respectively, the model produces highly accurate parameter estimates. Moreover, in scenarios where the sampling process for the FDD is influenced by the PAP $\mathbf{Z}$ (\textit{Scenarios 2} and \textit{3}), the model consistently provides reliable estimates for $\beta'(t)$, particularly when the FDD dimensionality is 100. However, a slight underestimation of $\beta'(t)$ is observed in cases where the FDD dimension exceeds 100.

For scenarios involving a strong PS effect dependent on $\mathbf{Y}$ (\textit{Scenarios 1} and \textit{3}), the model exhibits a tendency to underestimate $\beta(t)$. Despite this underestimation, the estimates remain statistically significant, underscoring the model’s capability to generate meaningful inferences even under challenging conditions.

\begin{table}
\caption{Median values (and 90\% confidence intervals) of relative bias of $\beta'(t)$ and $\beta(t)$ across sampling scenarios and combinations of samples’ dimensions $Comb(n^{I}(t),n^{D}(t))$.  Scenarios are defined as: \textit{Scenario 1} with $\beta'(t) = 0$ and $\beta(t) \sim N(2,0.25)$, \textit{Scenario 2} with $\beta'(t) \sim N(2,0.25)$ and $\beta(t) = 0$, and \textit{Scenario 3} with $\beta'(t) \sim N(1,0.25)$ and $\beta(t) \sim N(2,0.25)$ for $t=\{t_1,\cdots,t_4\}$.}
\label{tab:pref_results}
\centering
\small
\begin{tabular}{ccccc}
\hline
Parameter                                           &  \begin{tabular}[c]{@{}c@{}}Combination of\\ sample dimensions\end{tabular}  & \textit{Scenario 1} & \textit{Scenario 2} & \textit{Scenario 3} \\ \hline
\multicolumn{1}{c}{\multirow{4}{*}{$\beta'({t_1})$}} & $Comb(100,100)$ & -0.01 (-0.22,3.92) & -0.22 (-0.93,0.35) & -0.66 (-0.93,0.75) \\ 
                                                    & $Comb(100,200)$ & -0.01 (-0.13,1.75) & -0.22 (-0.92,0.41) & -0.97 (-1.02,-0.21) \\          
                                                    & $Comb(100,500)$ & 0.00 (-0.09,1.05)&  -0.32 (-0.88,0.00) & -0.96 (-0.99,-0.50) \\
                                                    & $Comb(200,100)$ & 0.04 (-0.26,2.31) & -0.09 (-0.58,0.42) & -0.95 (-1.04,0.91) \\ \hline
\multicolumn{1}{c}{\multirow{4}{*}{$\beta'({t_2})$}} & $Comb(100,100)$ & 0.38 (0.03,3.55) & -0.27 (-0.93,0.29) & -0.73 (-0.97,1.20) \\
                                                    & $Comb(100,200)$ & 0.35 (-0.08,1.99) & -0.31 (-0.89,0.21) & -1.03 (-1.10,0.07) \\
                                                    & $Comb(100,500)$ & 0.36 (0.21,1.16) & -0.41 (-0.89,-0.14) & -1.01 (-1.06,-0.29) \\
                                                    & $Comb(200,100)$ & 0.27 (-0.14,2.40) & -0.19 (-0.60,0.23) & -1.02 (-1.12,2.13) \\ \hline
\multicolumn{1}{c}{\multirow{4}{*}{$\beta'({t_3})$}} & $Comb(100,100)$ & -0.35 (-0.67,5.75) & -0.20 (-0.93,0.32) & -1.25 (-2.08,4.05) \\
                                                    & $Comb(100,200)$ & -0.31 (-0.43,2.88) & -0.20 (-0.90,0.42) & -0.78 (-1.09,9.95) \\
                                                    & $Comb(100,500)$ & -0.30 (-0.38,1.64) & -0.32 (-0.88,0.03) & -0.62 (-0.91,5.06) \\
                                                    & $Comb(200,100)$ & -0.32 (-0.79,1.86) & -0.15 (-0.54,0.38) & -0.49 (-1.02,24.02) \\ \hline
\multicolumn{1}{c}{\multirow{4}{*}{$\beta'({t_4})$}} & $Comb(100,100)$ & -0.12 (-0.65,6.70) & -0.30 (-0.94,0.18) & -2.36 (-4.61,9.41) \\ 
                                                    & $Comb(100,200)$ & -0.11 (-0.28,2.10) & -0.32 (-0.87,0.25) & -0.66 (-1.73,34.48) \\ 
                                                    & $Comb(100,500)$ & -0.09 (-0.18,1.63) & -0.42 (-0.90,-0.18) & -0.03 (-1.16,20.04) \\ 
                                                    & $Comb(200,100)$ & -0.14 (-0.79,3.13) & -0.19 (-0.54,0.11) & -0.32 (-2.04,79.84) \\ \hline           
\multicolumn{1}{c}{\multirow{4}{*}{$\beta({t_1})$}} & $Comb(100,100)$ & -0.27 (-0.47,-0.10) & 0.06 (-0.19,10.65) & -0.27 (-0.55,0.13) \\
                                                    & $Comb(100,200)$ & -0.36 (-0.49,-0.18) & 0.03 (-0.12,5.43) & -0.55 (-0.69,-0.42) \\
                                                    & $Comb(100,500)$ & -0.36 (-0.50,-0.26) & 0.04 (-0.08,3.51) & -0.58 (-0.70,-0.49) \\
                                                    & $Comb(200,100)$ & -0.24 (-0.42,-0.03) & 0.07 (-0.27,1.74) & -0.52 (-0.69,-0.36) \\ \hline
\multicolumn{1}{c}{\multirow{4}{*}{$\beta({t_2})$}} & $Comb(100,100)$ & -0.33 (-0.47,-0.22) & 0.11 (-0.25,14.98) & -0.34 (-0.54,-0.04) \\
                                                   & $Comb(100,200)$ & -0.40 (-0.49,-0.28) & 0.05 (-0.21,8.01) & -0.48 (-0.58,-0.34) \\                   
                                                   & $Comb(100,500)$ & -0.45 (-0.53,-0.36) & 0.04 (-0.11,4.84) & -0.52 (-0.62,-0.43) \\
                                                   & $Comb(200,100)$ & -0.32 (-0.45,-0.19) & 0.05 (-0.31,1.65) & -0.44 (-0.55,-0.27) \\ \hline
\multicolumn{1}{c}{\multirow{4}{*}{$\beta({t_3})$}} & $Comb(100,100)$ & -0.35 (-0.51,-0.17) & 0.08 (-0.18,9.73) & -0.34 (-0.48,-0.09) \\ 
                                                   & $Comb(100,200)$ & -0.40 (-0.55,-0.23) & 0.03 (-0.14,5.46) & -0.35 (-0.49,-0.17) \\
                                                   & $Comb(100,500)$ & -0.48 (-0.61,-0.37) & 0.05 (-0.07,3.43) & -0.42 (-0.57,-0.29) \\
                                                   & $Comb(200,100)$ & -0.33 (-0.50,-0.11) & 0.06 (-0.20,1.29) & -0.31 (-0.47,-0.07) \\ \hline
\multicolumn{1}{c}{\multirow{4}{*}{$\beta({t_4})$}} & $Comb(100,100)$ & -0.36 (-0.49,-0.21) & 0.11 (-0.21,15.04) & -0.29 (-0.42,-0.10) \\ 
                                                   & $Comb(100,200)$ & -0.42 (-0.53,-0.28) & 0.03 (-0.18,8.25) & -0.01 (-0.22,0.22) \\ 
                                                   & $Comb(100,500)$ & -0.50 (-0.61,-0.39) & 0.04 (-0.08,4.93) & -0.13 (-0.31,0.03) \\ 
                                                   & $Comb(200,100)$ & -0.34 (-0.49,-0.16) & 0.05 (-0.30,2.05) & 0.06 (-0.24,0.43) \\ \hline           
\end{tabular}
\end{table}

\normalsize

Expanding the sample size of the FDD emerges as a crucial factor in enhancing the stability and reducing the variability of parameter estimates. Conversely, increasing the sample size of the FID does not significantly affect the estimation of the preferential parameters. This finding aligns with theoretical expectations, as the estimation of $\beta(t)$ and $\beta'(t)$ relies exclusively on the information contained within the FDD. The absence of any observable impact from the FID underscores the distinct and non-overlapping contributions of FID and FDD.

\subsubsection{Evaluation of the estimation of spatio-temporal covariance parameters}

Distinct estimation patterns emerge for the range parameters across specific combinations of sample sizes and scenarios (Table \ref{tab:spat_param}). The parameter $\phi_{\mathbf{V}}$ is estimated reliably in most cases, with the exception of a slight overestimation observed under the combination $Comb(100,500)$ in \textit{Scenario 2}. In contrast, $\phi_{\mathbf{U}}$ is estimated accurately only in \textit{Scenario 2}, with overestimation occurring across the remaining scenarios. The joint model consistently provides accurate estimates of $\phi_{\mathbf{W}}$ across all tested combinations and scenarios.

Increasing sample sizes generally lead to greater uncertainty in the estimation of $\phi_{\mathbf{U}}$ and $\phi_{\mathbf{V}}$, particularly as the dimensionality of the FDD increases. This highlights the role of sample size and dimensionality in influencing the precision of range parameter estimates.

\begin{table}
\caption{Mode values (and 90\% confidence intervals) for $\phi_{\mathbf{V}}=0.15$, $\sigma_{\mathbf{V}}=0.80$, $\phi_{\mathbf{U}}=0.20$, $\sigma_{\mathbf{U}}=1.00$, $\phi_{\mathbf{W}}=0.20$, $\sigma_{\mathbf{W}}=1.00$, and $\delta=0.80$ across sampling scenarios and combinations of samples’ dimensions $Comb(n^{I}(t),n^{D}(t))$.  Scenarios are defined as: \textit{Scenario 1} with $\beta'(t) = 0$ and $\beta(t) \sim N(2,0.25)$, \textit{Scenario 2} with $\beta'(t) \sim N(2,0.25)$ and $\beta(t) = 0$, and \textit{Scenario 3} with $\beta'(t) \sim N(1,0.25)$ and $\beta(t) \sim N(2,0.25)$ for $t=\{t_1,\cdots,t_4\}$.}
\label{tab:spat_param}
\centering
\small
\begin{tabular}{ccccc}
\hline
Parameter                                                & \begin{tabular}[c]{@{}c@{}}Combination of\\ sample dimensions\end{tabular} & \textit{Scenario 1} & \textit{Scenario 2} & \textit{Scenario 3} \\ \hline
\multicolumn{1}{c}{\multirow{4}{*}{$\phi_{\mathbf{V}}$}} & $Comb(100,100)$ & 0.08 (0.02,0.58) & 0.35 (0.09,0.41) & 0.16 (0.07,0.43) \\
\multicolumn{1}{c}{}                                     & $Comb(100,200)$ & 0.10 (0.06,0.77) & 0.38 (0.13,0.47) & 0.10 (0.05,0.41) \\                                
\multicolumn{1}{c}{}                                     & $Comb(100,500)$ & 0.15 (0.09,1.06) & 0.48 (0.13,0.84) & 0.12 (0.08,0.61) \\
\multicolumn{1}{c}{}                                     & $Comb(200,100)$ & 0.15 (0.06,0.67) & 0.31 (0.22,0.41)  & 0.10 (0.06,0.51) \\ \hline
\multicolumn{1}{c}{\multirow{4}{*}{$\sigma_{\mathbf{V}}$}} &  $Comb(100,100)$ & 1.62 (0.00,3.10) & 0.89 (0.53,2.45) & 1.15 (0.36,2.22) \\
\multicolumn{1}{c}{}                                    & $Comb(100,200)$ & 1.37 (0.03,1.99) & 1.13 (0.63,1.95) & 1.48 (0.44,2.13) \\
\multicolumn{1}{c}{}                                    & $Comb(100,500)$ & 1.22 (0.36,2.01) & 1.78 (1.08,2.66) & 1.37 (0.53,2.10) \\
\multicolumn{1}{c}{}                                    & $Comb(200,100)$ & 1.25 (0.01,1.73) & 0.81 (0.54,1.34) & 1.37 (0.13,1.94) \\ \hline
\multicolumn{1}{c}{\multirow{4}{*}{$\phi_{\mathbf{U}}$}} & $Comb(100,100)$ & 0.32 (0.27,0.41) & 0.21 (0.15,0.42) & 0.30 (0.25,0.42) \\
\multicolumn{1}{c}{}                                    & $Comb(100,200)$ & 0.35 (0.31,0.44) & 0.22 (0.15,0.43) & 0.37 (0.31,0.46) \\
\multicolumn{1}{c}{}                                    & $Comb(100,500)$ & 0.46 (0.40,0.67) & 0.21 (0.15,0.62) & 0.48 (0.38,0.70) \\
\multicolumn{1}{c}{}                                    & $Comb(200,100)$ & 0.31 (0.25,0.39) & 0.22 (0.18,0.38) & 0.29 (0.24,0.38) \\ \hline
\multicolumn{1}{c}{\multirow{4}{*}{$\sigma_{\mathbf{U}}$}} & $Comb(100,100)$ & 1.25 (0.90,1.41) & 0.94 (0.06,1.17) & 1.08 (0.87,1.39) \\
\multicolumn{1}{c}{}                                    & $Comb(100,200)$ & 1.43 (1.13,1.67) & 1.01 (0.20,1.19) & 1.30 (1.01,1.51) \\
\multicolumn{1}{c}{}                                    & $Comb(100,500)$ & 1.79 (1.46,2.44) & 1.00 (0.37,1.19) & 1.65 (1.35,2.40) \\
\multicolumn{1}{c}{}                                    & $Comb(200,100)$ & 1.17 (0.96,1.39) & 0.98 (0.33,1.17) & 1.17 (0.90,1.33) \\ \hline
\multicolumn{1}{c}{\multirow{4}{*}{$\phi_{\mathbf{W}}$}} & $Comb(100,100)$ & 0.22 (0.17,0.28) & 0.20 (0.17,0.28) & 0.23 (0.19,0.30) \\
\multicolumn{1}{c}{}                                    & $Comb(100,200)$ & 0.20 (0.17,0.25) & 0.20 (0.16,0.25) & 0.21 (0.19,0.27) \\
\multicolumn{1}{c}{}                                    & $Comb(100,500)$ & 0.20 (0.18,0.25) & 0.21 (0.18,0.25) & 0.23 (0.20,0.26) \\
\multicolumn{1}{c}{}                                    & $Comb(200,100)$ & 0.23 (0.18,0.29) & 0.24 (0.18,0.26) & 0.21 (0.18,0.27) \\ \hline
\multicolumn{1}{c}{\multirow{4}{*}{$\sigma_{\mathbf{W}}$}} & $Comb(100,100)$ & 0.93 (0.83,1.14) & 0.98 (0.83,1.38) & 0.95 (0.81,1.14) \\
\multicolumn{1}{c}{}                                    & $Comb(100,200)$ & 1.02 (0.89,1.17) & 1.05 (0.91,1.35) & 1.03 (0.91,1.13) \\
\multicolumn{1}{c}{}                                    & $Comb(100,500)$ & 1.06 (0.95,1.21) & 1.04 (1.02,1.41) & 1.04 (0.95,1.16) \\
\multicolumn{1}{c}{}                                    & $Comb(200,100)$ & 0.96 (0.84,1.11) & 0.98 (0.85,1.27) & 0.94 (0.86,1.12) \\ \hline
\multicolumn{1}{c}{\multirow{4}{*}{$\delta$}}           & $Comb(100,100)$ & 0.82 (0.69,0.88) & 0.84 (0.72,0.91) & 0.83 (0.72,0.88) \\ 
\multicolumn{1}{c}{}                                    & $Comb(100,200)$ & 0.83 (0.73,0.86) & 0.81 (0.72,0.90) & 0.83 (0.70,0.87) \\ 
\multicolumn{1}{c}{}                                    & $Comb(100,500)$ & 0.83 (0.75,0.87) & 0.83 (0.76,0.90) & 0.83 (0.76,0.87) \\                 
\multicolumn{1}{c}{}                                    & $Comb(200,100)$ & 0.80 (0.72,0.87) & 0.81 (0.72,0.89) & 0.78 (0.70,0.87) \\ \hline        
\end{tabular}
\end{table}

\normalsize

For most evaluated model configurations, marginal standard deviations are estimated with high accuracy. Exceptions are observed for $\sigma_{\mathbf{U}}$ in \textit{Scenarios 1} and \textit{3}, particularly in cases where the FDD dimension exceeds the FID dimension. Additionally, uncertainty in parameter estimation generally decreases as the FDD sample size increases. Among the latent fields, the model consistently achieves the highest accuracy in estimating parameters related to the field $\mathbf{W}$.

The temporal correlation parameter $\delta$ is accurately estimated across all model configurations. Similarly, an increase in sample size reduces the uncertainty associated with its estimation, reflecting improved precision under larger sample dimensions.

\subsubsection{Evaluation of the estimation of intercept and regression coefficient parameters}

The proposed model provides accurate estimates for $\alpha$ and $\alpha'$, as demonstrated by the inclusion of the true value (represented by the \textcolor{red}{red} points in Figure 2 of the Supplementary Material \citep{Silva2025}) within the range of estimated values across all model configurations (Figures 2a) and 2b)) of the Supplementary Material \citep{Silva2025}). However, the accuracy of these estimates tends to be higher for smaller sample dimensions, with the exception of $\alpha$ under \textit{Scenario 2}.

In contrast, the intercept parameter associated with the point process, denoted as $\alpha''(t)$, exhibits a tendency for overestimation (Figure 3 of the Supplementary Material \citep{Silva2025}). This overestimation becomes more pronounced as the FDD dimension increases. Based on the definition of the intensity process in \eqref{eq:ipp}, the expected value of the point process intensity related to the FDD, conditioned on the zero-mean GFs $\mathbf{V}$ and $\mathbf{U}$, is represented by $\alpha''(t)$. According to the properties of the intensity function, this parameter is inherently positive and increases with the sample size. Consequently, this behavior explains why $\alpha''_t$ is consistently estimated above its true value (which is defined as zero) and why the bias increases with the dimensionality of the FDD.

\begin{figure}[!h]
    \centering
    \begin{subfigure}{0.9\linewidth}
        \includegraphics[width=0.5\linewidth]{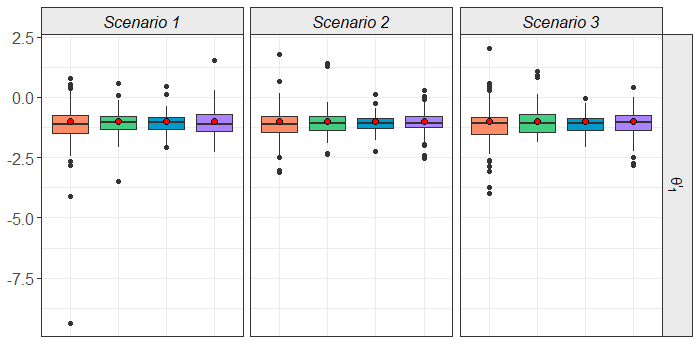}\qquad
        \includegraphics[width=0.5\linewidth]{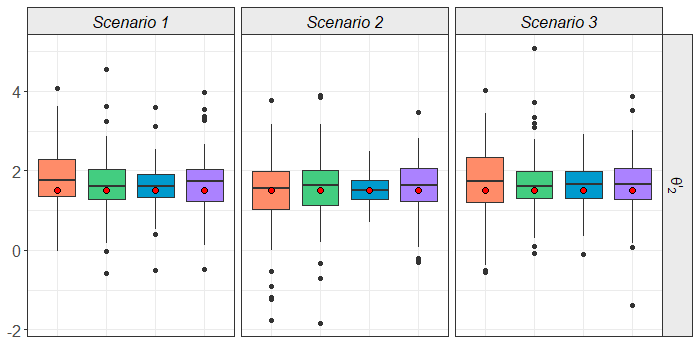}\\
        \includegraphics[width=0.5\linewidth]{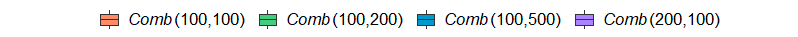}\qquad
        \includegraphics[width=0.5\linewidth]{combinations_legend.png}
        \caption{$\theta_j'$}
        \label{fig:reg_z}
    \end{subfigure}\\
    \begin{subfigure}{0.9\linewidth}
        \includegraphics[width=0.5\linewidth]{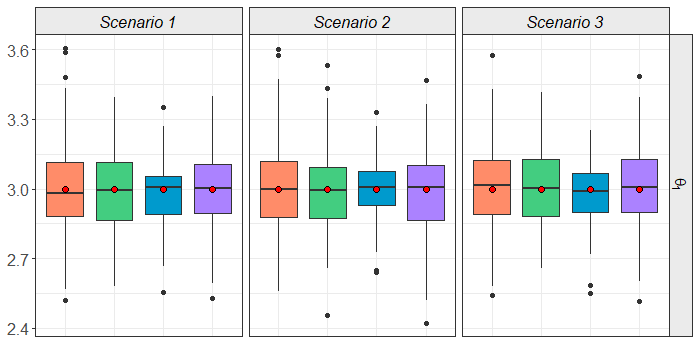}\qquad
        \includegraphics[width=0.5\linewidth]{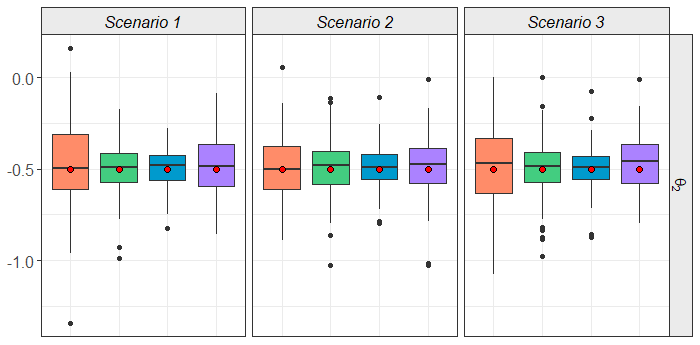}\\
        \includegraphics[width=0.5\linewidth]{combinations_legend.png}\qquad
        \includegraphics[width=0.5\linewidth]{combinations_legend.png}
        \caption{$\theta_j$}
        \label{fig:reg_y}
    \end{subfigure}
    \caption{Estimates of regression coefficients (with true values $\mathbf{\theta'}=(\theta_1'=-1$, $\theta_2'=1.5$) and $\mathbf{\theta}=(\theta_1=1$,$\theta_2=-0.5$))  and combinations of samples’ dimensions $Comb(n^{I}(t),n^{D}(t))$.  Scenarios are defined as: \textit{Scenario 1} with $\beta'(t) = 0$ and $\beta(t) \sim N(2,0.25)$, \textit{Scenario 2} with $\beta'(t) \sim N(2,0.25)$ and $\beta(t) = 0$, and \textit{Scenario 3} with $\beta'(t) \sim N(1,0.25)$ and $\beta(t) \sim N(2,0.25)$ for $t=\{t_1,\cdots,t_4\}$. The \textcolor{red}{red} points represent the true values of the corresponding parameter for all 100 replicas.}
    \label{fig:reg_coef_by_comb_dim}
\end{figure}

Regarding the estimation of the regression coefficients, the model demonstrates accuracy across all configurations, as the true values (depicted by the \textcolor{red}{red} points in Figure \ref{fig:reg_coef_by_comb_dim}) consistently fall within the interquartile range and are closely aligned with the median value. Additionally, the data dimension plays a significant role in the precision of the estimates, with estimation accuracy improving as the data dimension increases.

\subsubsection{Evaluation of the prediction performance}

In terms of predictive performance (Figure 4 of the Supplementary Material \citep{Silva2025} and Figure \ref{fig:model_perf}), no significant differences are observed across the various model configurations, although a slight improvement in prediction accuracy is evident with larger datasets. An analysis of the Hellinger distances estimates across different scenarios (Figure \ref{fig:model_perf}) demonstrates its ability to generate predicted data distributions that closely match the observed distributions, as indicated by lower values of the Hellinger distance estimates.

\begin{figure}[!h]
    \centering
        \includegraphics[width=0.6\linewidth]{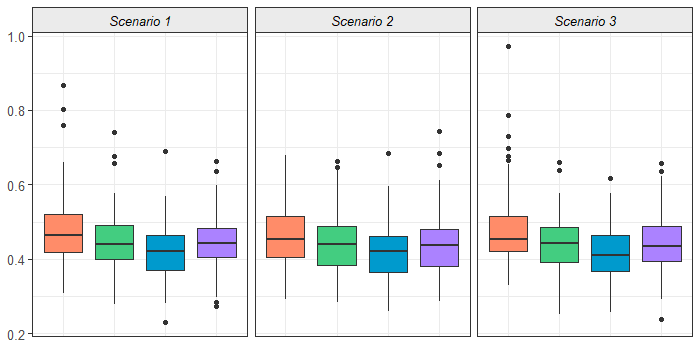}\\
        \includegraphics[width=0.6\linewidth]{combinations_legend.png}
    \caption{Evaluation of predictive performance, using the Hellinger distance, across sampling scenarios and combinations of samples' dimensions $Comb(n^{I}(t),n^{D}(t))$. Scenarios are defined as: \textit{Scenario 1} with $\beta'(t) = 0$ and $\beta(t) \sim N(2,0.25)$, \textit{Scenario 2} with $\beta'(t) \sim N(2,0.25)$ and $\beta(t) = 0$, and \textit{Scenario 3} with $\beta'(t) \sim N(1,0.25)$ and $\beta(t) \sim N(2,0.25)$ for $t=\{t_1,\cdots,t_4\}$.}
    \label{fig:model_perf}
\end{figure}

\subsubsection{Evaluation of the contribution of each data source}

A comparison of the predictive performance (RMSE and MAE) across the three models reveals that the joint model outperforms the individual models (Figure 5 of the Supplementary Material \citep{Silva2025}). This advantage is most evident when evaluating the Hellinger distance, as the joint model consistently generates predicted distributions that more closely resemble the true distributions (Figure \ref{fig:model_perf_3models}). Additionally, the FID model demonstrates the poorest predictive performance, with the exception of the model configuration $C_n(200,100)$, where the FID model benefits from the higher dimension of FID. The performance of the FDD model closely approximates that of the joint model when the FDD dimension significantly exceeds the FID dimension.

A common trend observed across all models is the impact of sample dimension: as the sample dimension increases, predictive performance improves.

\begin{figure}[!h]
 \centering
        \includegraphics[width=0.6\linewidth]{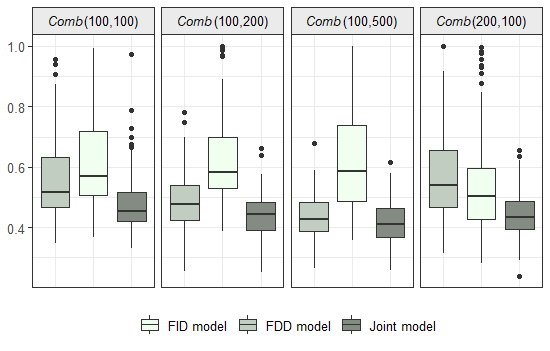}
    \caption{Evaluation of the contribution of each data source, using the Hellinger distance, across model configurations of sample dimension $Comb(n^{I}(t),n^{D}(t))$ under \textit{Scenario 3} ($\beta'(t)\sim N(1,0.25)$ and $\beta(t)\sim N(2,0.25)$) for $t=\{t_1,\cdots,t_4\}$.}
    \label{fig:model_perf_3models}
\end{figure}

\section{Application to sardine data}

Given the socioeconomic importance of small pelagic fish for the fisheries of several countries worldwide, and the important effect of global change on their spatial distribution, we undertake the task of predicting its spatio-temporal distribution within the southern Portuguese shelf, using the Iberian sardine as a case study, for which a large dataset of FID and FDD data is available.

\subsection{Data}

\subsubsection{Fishery-independent data}

\begin{figure}[!h]
    \centering
    \includegraphics[scale=0.6]{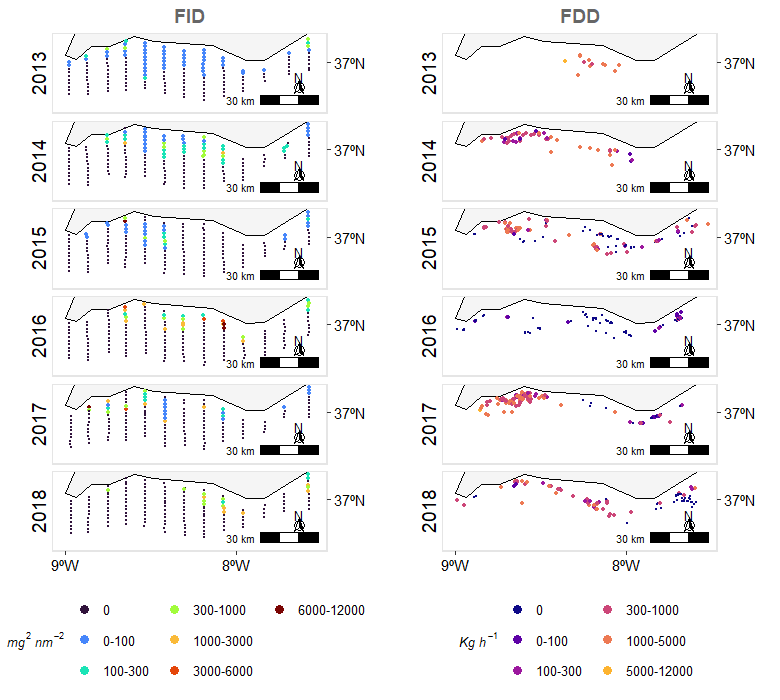}
   \caption{Observed biomass index of sardine off the southern coast of Portugal between 2013 and 2018 from FDD source in $mg^2~nm^{-2}$ (first column) and from FDD source in $Kg~h^{-1}$ (second column).}
    \label{fig:observ_biomass}
\end{figure}

The spatio-temporal distribution of sardine biomass is assessed using data from the Portuguese spring acoustic (PELAGO) series (first column of Figure \ref{fig:observ_biomass}), conducted by the Portuguese Institute for the Sea and Atmosphere (IPMA) in Portuguese continental waters between 2013 and 2018.

\begin{table}[h]
\centering
\caption{Summary of recorded locations of sardine off the southern coast of Portugal, along with the count and percentage of locations exhibiting strictly positive values (i.e., presences), for both data sources FID (PELAGO survey series data) and FDD (commercial data obtained through the AIS) between 2013 and 2018. The table includes the overall estimate of the nautical area-scattering coefficient (NASC) derived from FID and the total Catch Per Unit Effort (CPUE) from FDD for sardine.}
\small
\begin{tabular}{ccccr}
\hline
\multicolumn{1}{c}{Data source} & \multicolumn{1}{c}{Year} & \begin{tabular}[c]{@{}c@{}}Number of\\ locations\end{tabular} & \begin{tabular}[c]{@{}c@{}}Sardine positive\\ locations\end{tabular} & \multicolumn{1}{c}{\begin{tabular}[c]{@{}c@{}}Total \\ estimated/captured\end{tabular}} \\ \hline
\multirow{6}{*}{FID}               & 2013 & 147 & 54 (37\%) & 5090.70 $mg^2~nm^{-2}$ \\
                                   & 2014 & 150 & 48 (32\%) & 12056.10 $mg^2~nm^{-2}$ \\
                                   & 2015 & 152 & 31 (20\%) & 10791.30 $mg^2~nm^{-2}$ \\
                                   & 2016 & 162 & 26 (16\%) & 51415.60 $mg^2~nm^{-2}$ \\
                                   & 2017 & 144 & 29 (20\%) & 26142.06 $mg^2~nm^{-2}$ \\
                                   & 2018 & 147 & 17 (12\%) & 17852.80 $mg^2~nm^{-2}$ \\ \hline
\multirow{6}{*}{FDD}               & 2013 & 12 & 12 (100\%) & 33094.37 $Kg~h^{-1}$ \\
                                   & 2014 & 50 & 50 (100\%) & 43286.51 $Kg~h^{-1}$ \\
                                   & 2015 & 106 & 65 (61\%) & 58602.14 $Kg~h^{-1}$ \\
                                   & 2016 & 84 & 18 (21\%) & 5244.82 $Kg~h^{-1}$ \\
                                   & 2017 & 140 & 116 (83\%) & 150688.12 $Kg~h^{-1}$ \\
                                   & 2018 & 85 & 48 (56\%) & 39227.74 $Kg~h^{-1}$ \\ \hline
\end{tabular}
\label{tab:data}
\end{table}
\normalsize

The primary objective of the PELAGO survey series is to monitor the abundance, biomass and spatial distribution of adult sardines and anchovies, using this data in the assessment of stock status to support management. The survey design involves continuous daytime acoustic measurements along parallel transects, facilitated by a calibrated 38-kHz echosounder. Data processing includes integrating and averaging the resulting backscatter from the water column over 1 $nm$ intervals, expressed as nautical area-scattering coefficients [NASC; $S_A$ (in $m^2~nm^{-2}$)]. The inter-transect distance is consistently 6 $nm$. The detailed methodology underpinning the PELAGO series is outlined in \cite{Doray2021}.

Each NASC value, representing a proportion of fish density, is utilized as a biomass proxy for each pair of coordinates (longitude and latitude). The FID source incorporates 902 sardine NASC values recorded between 2013 and 2018, where the annual species absences vary between 63\% and 88\% (Table \ref{tab:data}).

\subsubsection{Fishery-dependent data}

For the same area of interest, the FDD source consists of recent output data by \cite{Araujo2023} generated from Automatic Identification System (AIS) data obtained under a commercial license for the Portuguese mainland purse seiners. Commercial data for each year aligns with the period when the scientific survey was conducted in the same year, ensuring consistency in the temporal scope to avoid variations in species distribution patterns that may occur throughout each year. The dataset from commercial source is standardized by fishing effort (total duration in hours of the fishing event), quantified in kilograms per hour ($Kg~h^{-1}$), enhancing comparison across samples. The FDD dataset comprises 477 commercial samples, providing valuable insights into the spatial distribution of sardine biomass in the studied region (second column of Figure \ref{fig:observ_biomass}). Conversely, the majority of the FDD observations indicate species presence, varying between 21\% and 100\% (Table \ref{tab:data}).

\subsubsection{Covariates}

Since it provides a detailed and extensive analysis, we explore the covariate effects described in \cite{Silva2024} to investigate the relationship between sardine distribution and lagged environmental variables (Table \ref{tab:model_expr}). For example, 
$K(CHL,2,2)$ represents a weighted average of chlorophyll-a concentration measured on the day of biomass estimation and four days prior, with greater weight assigned to observations from two days before.

The dataset of daily environmental information was obtained from the COPERNICUS server (\url{https://resources.marine.copernicus.eu/products}), covering {the} {stu}{-dy} region and time frame. This dataset includes satellite-derived sea surface temperature (SST) in degrees Celsius (ESA SST CCI and C3S reprocessed SST analyses, \url{https://doi.org/10.48670/moi-00169}), satellite-derived chlorophyll-a concentration in $mg~m^{-3}$ (Global Ocean Colour project, \url{https://doi.org/10.48670/moi-00281}), bathymetry in meters, and the intensity and direction of ocean currents, with intensity measured in $m~s^{-1}$ and direction in degrees (Atlantic-Iberian Biscay Irish-Ocean Physics Reanalysis, \url{https://doi.org/10.48670/moi-00029}).

\begin{table}[h]
\caption{Model expression based on \cite{Silva2024} for studying the spatio-temporal distribution of sardine. Specific covariates are indicated by acronyms (SST: sea surface temperature, CHL: chlorophyll-a concentration, INT: intensity of ocean currents).}
\label{tab:model_expr}
\centering
\small
\begin{tabular}{cl}
\hline
\multicolumn{1}{l}{Process} & Expression                                                                                                                                           \\ \hline
$\mathbf{Z}$                & $\alpha' + K(CHL,2,2) + Bathymetry + K(INT,24,4) + \mathbf{V} + \mathbf{W} $ \\ \hline
$\mathbf{Y}$                & $\alpha + K(SST,0,0) + K(CHL,17,0) + Bathymetry + K(INT,10,6) + \mathbf{U} + \mathbf{W} $                                                    \\ \hline
\end{tabular}
\end{table}

\normalsize

Besides environmental variables, commercial vessel characteristics were used to evaluate their potential effects on catchability efficiency. These attributes include vessel length (in meters) and main engine power (in kilowatts), which were collected from the EU Fleet Register (\url{https://webgate.ec.europa.eu/fleet-europa/search_en}).  

\subsection{Catchability effect}

Given the measurement discrepancies between the two data sources (NASC for FID and $Kg~h^{-1}$ for FDD), and the fact that the FDD was collected through the commercial activity of multiple vessels, the relative biomass process $\mathbf{S^*}$ is defined as the process of primary interest. Specifically, survey data were collected using comparable research vessels, while commercial data were obtained from eighteen distinct fishing vessels. Therefore, the index $v$ takes values $v=\{1,\cdots,19\}$, where $v=1$ corresponds to the research vessel, and $v=\{2,\cdots,19\}$ correspond to the commercial vessels. Given the consistent use of the same sampling methodology for all acoustic surveys and the inherent differences between the survey and commercial vessels, $k(1)$ is treated as a constant while catchability for the remaining vessels is modeling through \eqref{eq:catch_effect}.

\subsection{Results}

By integrating FDD and FID, the proposed joint model provided valuable insights into the spatio-temporal distribution of sardines along the southern coast of the Portuguese shelf.

\begin{figure}[!h]
    \centering
    \includegraphics[scale=0.52]{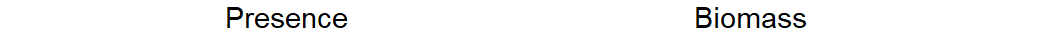}\\
    \includegraphics[scale=0.5]{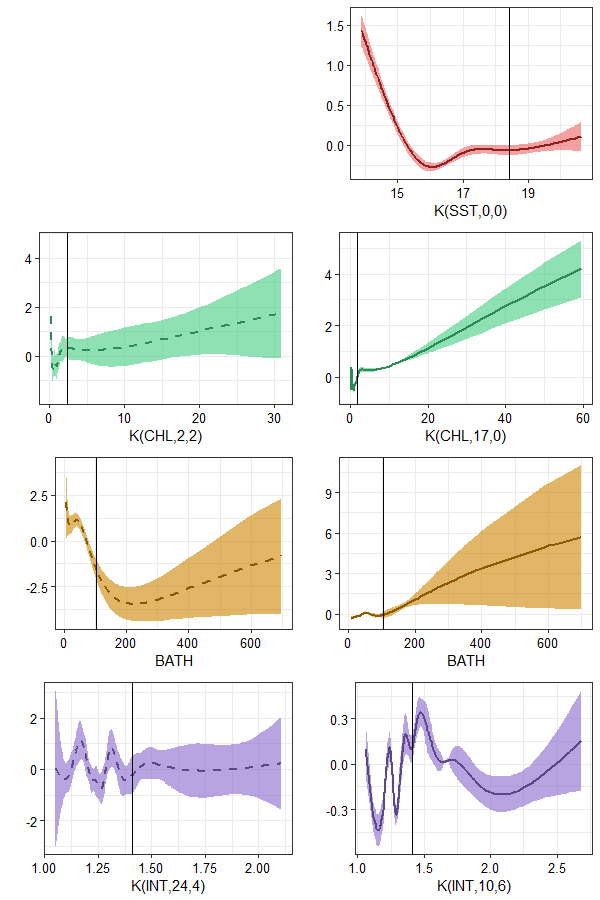}
   \caption{Fixed (environmental) effects for sardine presence (first column) and relative biomass (second column) derived from both FID (fishery-independent data) and FDD (fishery-dependent data) sources collected between 2013 and 2018. Certain covariates are represented by acronyms (SST: sea surface temperature, CHL: chlorophyll-a concentration, INT: intensity of ocean currents). Vertical lines depict the 80\% quantiles for each observed covariate, and $K(.)$ refers to the weighted average function of covariates defined in \cite{Silva2024}.}
    \label{fig:smooth_effects}
\end{figure}

The model estimated the fixed effects (Figure \ref{fig:smooth_effects}), highlighting the impact of environmental variables on sardine distribution. Specifically, bathymetry, chlorophyll-a concentration, and ocean current intensity were evaluated as explanatory variables for both sardine presence and the relative biomass index, as described by \citet{Silva2024}. Thus, SST was used solely to explain variations in the species' relative biomass index, emphasizing its unique role in driving biomass fluctuations.

The influence of chlorophyll-a concentration on sardine presence demonstrates a non-linear relationship, with a decreasing probability at lower concentrations and an increasing probability at higher values, suggesting an optimal range for sardine distribution. Sardine presence probability generally declines with increasing depth, reaching a minimum at 225 $m$, although a partial peak is observed at 40 $m$, indicating a depth preference for sardines in shallower waters. Furthermore, sardine presence is positively correlated with lower ocean current intensities, peaking at approximately 1.17 and 1.31 $m~s^{-1}$. However, beyond 1.33 $m~s^{-1}$, the effect becomes negligible, suggesting that sardines favor moderate current conditions.

For the sardine relative biomass index, a relevant relationship with SST is observed, characterized by a sharp decreasing trend from lower temperatures (around 14C) to 16ºC, followed by a smooth increase in biomass with higher temperatures. However, this increase is not statistically significant beyond approximately 18.5ºC. The relative biomass index also shows a relationship with the chlorophyll-a concentration, remaining stable up to 1 $mg~m^{-3}$ and then exhibiting a significant increase beyond this threshold. Additionally, sardine relative biomass increases with depth, indicating a positive association. Ocean current intensity exerts an oscillatory effect on relative biomass, peaking around 1.47 $m~s^{-1}$. At higher current intensities, its influence diminishes and becomes negligible.

Incorporating vessel characteristics into the catchability effects, the final model - selected based on the lowest Akaike Information Criterion (AIC) value - utilizes a single parameter to represent each vessel, excluding individual vessel-specific features (see Section 4 of the Supplementary Material, \citet{Silva2025}).

\begin{figure}[!h]
\centering
    \includegraphics[scale=0.60]{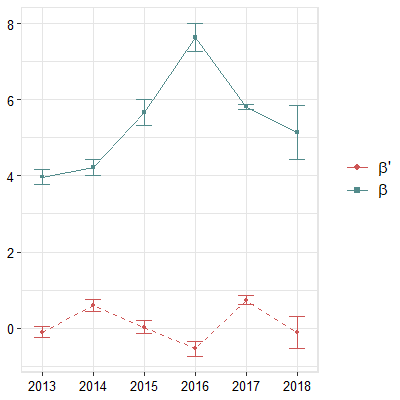}
   \caption{Annual estimates of preferential parameters associated with sardine \textcolor{indianred3}{\textbf{presence}} and \textcolor{darkslategray4}{\textbf{biomass}} derived from FDD (fishery-dependent data) source collected between 2013 and 2018 at the south coast of Portugal.}
    \label{fig:pref_par}
\end{figure}

The findings reveal a strong dependence of the FDD sampling process on both sardine presence and relative biomass. Specifically, the correlation between the spatial distribution of FDD data and the PAP is significantly negative in 2016 but positive in 2014 and 2017. Additionally, a consistently positive relationship is observed between the FDD sampling process and relative biomass across all years (Figure \ref{fig:pref_par}). This relationship is further demonstrated in Figure \ref{fig:observ_biomass}, where most FDD samples overlap with FID observations at locations corresponding to higher biomass values and sardine presence. In 2014 and 2017, FDD samples predominantly align with positive FID observations, coinciding with periods of elevated biomass presence.

The model identified distinct spatial dependence structures for the $\mathbf{Z}$ process (PAP) and the $\mathbf{Y}^*$ process (relative biomass, and consequently the biomass process $\mathbf{Y}$) (Table~\ref{tab:res_par}). Spatial autocorrelation for the PAP process $\mathbf{Z}$ decreases beyond approximately 19 km, whereas for the biomass process $\mathbf{Y}$, it persists up to 55 km. The shared spatio-temporal latent field exhibits a spatial range of approximately 8 km, with a mean annual dependence of 0.17, closely aligning with the estimate reported by \cite{Silva2024}.

\begin{table}[h]
\caption{Parameter estimates of covariance (and standard errors). Standard errors for $\sigma_.$, $\phi_.$ and $\delta$ are not provided since they resulted from the reparameterization of $\kappa_.$, $\tau_.$ and $\delta^{*}$ (see Section \ref{ssec:inf_est}).}
\centering
\small
\begin{tabular}{lr}
\hline
\multicolumn{1}{c}{Parameter} & Estimative (Std. Dev.) \\ \hline

$\phi_{\mathbf{V}}$ (Km) & 19.41\\
$\sigma_{\mathbf{V}}$ & 1.89\\
$log(\kappa_{\mathbf{V}})$ & -1.93 (0.10)\\
$log(\tau_{\mathbf{V}})$ & 0.02 (0.19)\\
$\phi_{\mathbf{U}}$ (Km) & 54.57\\
$\sigma_{\mathbf{U}}$ & 0.71\\
$log(\kappa_{\mathbf{U}})$ & -2.96 (0.06)\\
$log(\tau_{\mathbf{U}})$ & 2.04 (0.12)\\
$\phi_{\mathbf{W}}$ (Km) & 8.00\\
$\sigma_{\mathbf{W}}$ & 1.40\\
$log(\kappa_{\mathbf{W}})$ & -1.04 (0.03)\\
$log(\tau_{\mathbf{W}})$ & -0.56 (0.06)\\
$\delta$ & 0.17\\
$\delta^{*}$ & 0.34 (0.05)\\   \hline
\end{tabular}
\label{tab:res_par}
\end{table}

\normalsize

\begin{figure}[!h]
    \centering
    \includegraphics[scale=0.5]{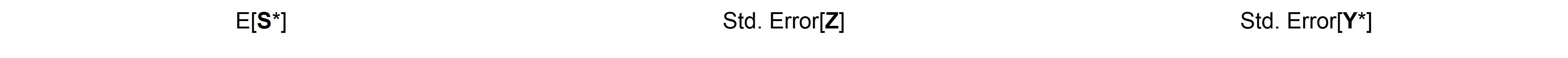}\\
    \includegraphics[scale=0.43]{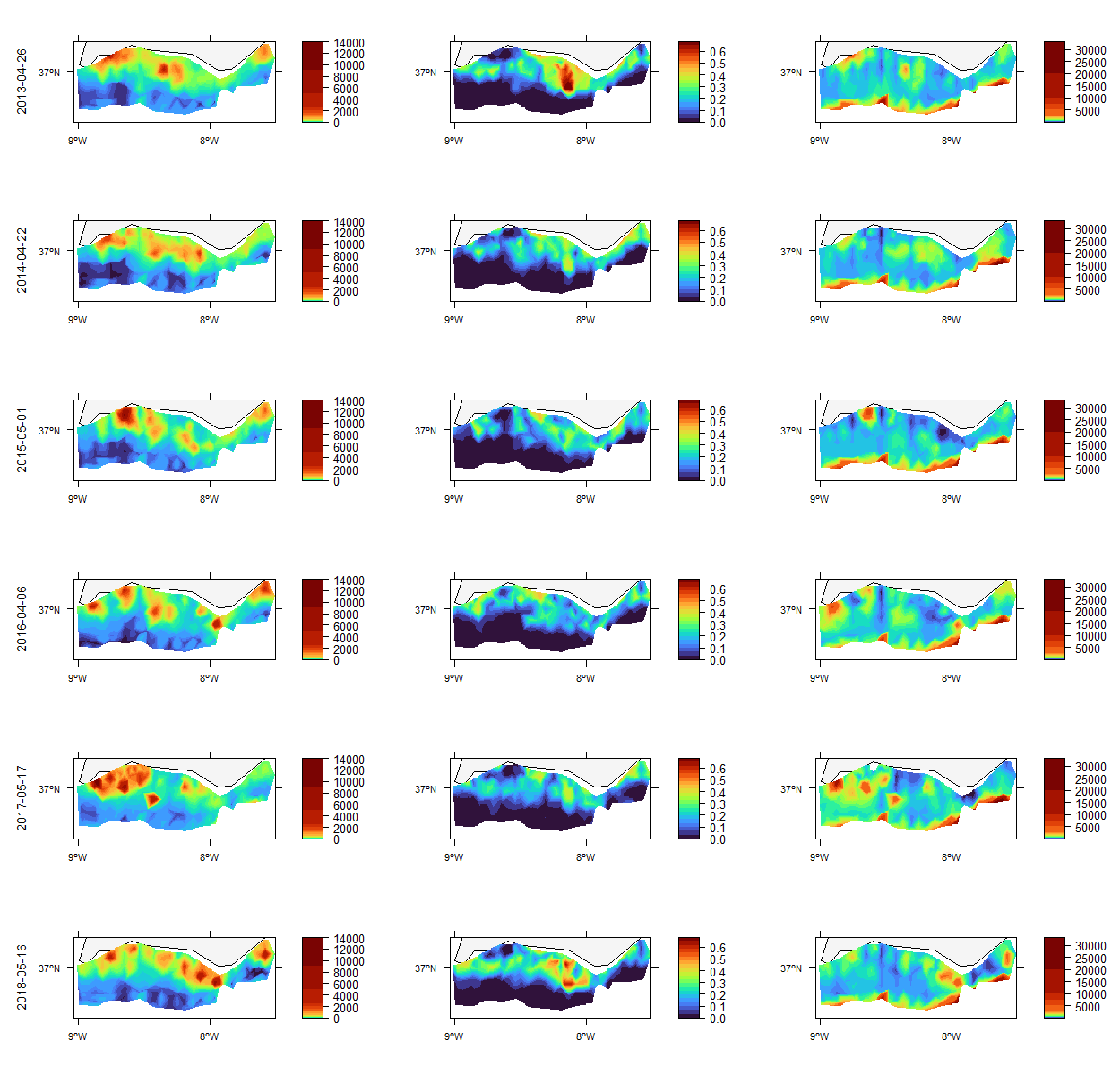}
   \caption{Predicted relative biomass index of sardines, $E[\mathbf{S^*}]=\hat{\pi}\cdot\hat{\mu}$, in the Portuguese south coast (left column), along with standard errors associated with the predicted probability of presence $\hat{\pi}$ (middle column) and the predicted relative biomass under presence $\hat{\mu}$ (right column). Each row presents the results for the representative day of each year in the study period (2013-2018).}
    \label{fig:pred_biomass}
\end{figure}

The modeling process enabled the mapping of predicted relative biomass for each year (Figure \ref{fig:pred_biomass}), with predictions corresponding to the central days of each annual period analyzed. Distinct spatial distribution patterns of sardine relative biomass emerged, revealing a general decrease in relative biomass with increasing distance from the coast. Although the locations of sardine hotspots varied across years, several persistent coastal hotspots were identified throughout the study period, particularly along the western coastal regions.

\section{Discussion}

The integration of data from diverse sampling processes poses substantial challenges in marine ecology, particularly for accurately describing fish distributions. Both FID and FDD provide valuable but distinct insights into fish distribution \citep{Pennino2016, Izquierdo2022, Silva2024}. These differences in sampling methodologies necessitate robust joint modeling frameworks capable of accounting for their unique characteristics.To address this complexity, we formulated and programmed a novel joint spatio-temporal model to infer fish distribution patterns combining FID and FDD data. 

Our approach extends previous methodologies by explicitly modeling ZI data, disentangling the effects of PAP and relative biomass on FDD sampling processes, and estimating the vessel catchability efficiency. This methodological advancement addresses critical limitations in single-source models and improves predictive accuracy, particularly in the presence of PS.

Building on this foundation, the model provides quantitative insights into fishing behavior through the PS parameters. These coefficients characterize how fishing effort responds to species presence and biomass, enabling the assessment of spatial selectivity and effort allocation. By estimating separate PS effects for PAP and relative biomass, the model offers distinct advantages over approaches in \cite{Rufener2021} and \cite{Alglave2022}, which do not explicitly account for this separation. This information is critical for understanding patterns of fishing pressure and supports spatially explicit management strategies, such as identifying areas of high exploitation or evaluating the effectiveness of spatial regulations. 

Simulation studies demonstrate the model's robust performance across diverse scenarios, producing unbiased estimates for most parameters. Specifically, the model accurately estimated the time correlation parameter $\delta$, the intercept parameters ($\alpha$ and $\alpha'$) of the linear predictors, and the regression parameters ($\theta$ and $\theta'$). Additionally, the predictive performance of the joint model consistently surpasses that of individual FID and FDD models, underscoring the value of integrating heterogeneous data sources for spatio-temporal modeling.

Nonetheless, some limitations were identified, mainly related to the estimation of the PS and $\alpha''(t)$ parameters. In particular, slight biases in the $\beta'(t)$ and $\beta(t)$ estimates were obtained, echoing findings by \citet{Silva2024_SPE}. However, the model significantly captures varying degrees of PS and their signals, even under scenarios with strong preferentiality. The reduction of variability in parameter estimates with increasing FDD sample sizes aligns with established statistical principles, emphasizing the utility of larger datasets in ecological modeling. The overestimation of the intercept for the intensity function, $\alpha''(t)$, aligns with theoretical expectations, given that it represents the mean of the intensity process and its consequent positive relationship with sample size. Despite this overestimation does not exhibit a clear or consistent relationship with biases in the estimation of the PS parameters, but $\alpha''(t)$ tends to be more heavily overestimated when $\beta(t)$ is underestimated, suggesting a compensatory mechanism within the model that aims to maintain accurate point process intensity. This interaction may play a critical role in preserving the quality of spatial predictions, even when some parameter estimates are biased.

The real-world application to sardine populations along the Portuguese continental shelf further highlights the model's capability to integrate environmental covariates and vessel-specific features. The observed relationships between relative biomass and environmental variables, such as SST, chlorophyll-a concentration, and ocean currents, corroborate previous findings in \citet{Silva2024}. However, the inclusion of FDD refines effect estimates and reveals previously undetected patterns, such as the significant influence of ocean current direction on sardine biomass. The most notable difference arises in the spatial distribution predictions, where the current study reveals more clearly defined hotspots, likely due to the commercial nature of FDD, which targets areas of higher sardine abundance.

The principal strength of our model lies in its ability to integrate heterogeneous data sources while accounting for PS, enhancing the precision and ecological relevance of species distribution models. Nonetheless, limitations remain - particularly when strong preferential effects simultaneously influence both PAP and biomass components, where estimation becomes less reliable.

Future research should extend this approach to incorporate a distinction between juvenile and adult sardine populations. Modeling these groups separately could reveal patterns influenced by fishing regulations or market preferences, particularly in juvenile aggregation zones. Exploring how these factors interact with FID and FDD sampling processes will enhance the ecological and management relevance of joint models in marine studies. Additional future work could explore extensions of the model by integrating alternative covariance structures (e.g., exponential or Gaussian) or longer temporal dependencies through higher-order autoregressive processes.

Finally, the capability of the proposed model to integrate diverse data sources, including external covariates, positions it as a powerful tool for a variety of disciplines when data sources with these sampling schemes are available. Some areas may include epidemiology and public health.

\section*{Acknowledgments}
The authors would like thanks the teams that collected the data. The biological survey datasets generated during and/or analyzed during the current study are available from the Portuguese Data Collection Framework on reasonable request. It is not ethically feasible to share any AIS data, as it would publicly reveal vessel information indicating where the activity takes place, while disclosing sensitive information. Additionally, the AIS data outputs are ruled by a confidentiality agreement between the different authors, preventing the share of the provided AIS data outputs, any private information regarding the fishing vessel and other related information.


\section*{Funding}
This study received support from the SARDINHA2030 project (MAR-111.4.1-FEAMPA-00001), and the project UID/00013 through the Centre for Mathematics of University of Minho (CMAT/UM).

\newpage

\bibliographystyle{apalike}
\bibliography{Manuscript}

\newpage

\appendix
\counterwithin{equation}{section}
\counterwithin{figure}{section}
\section{Likelihood of the model}
\label{sec:theory}

\subsection{Likelihood for $\mathbf{Y} \vert \mathbf{X},\mathbf{X}^{S},\mathbf{U},\mathbf{W}$}
\small
\begin{align}
\mathcal{L}(\zeta, \sigma; \mathbf{y}) = & \prod_{i=1}^{n} \frac{y_i^{\left( \frac{\zeta^2}{\sigma^2}-1 \right)} e^{-\left(\frac{\zeta}{\sigma^2}\right)y_i} \left( \frac{\zeta}{\sigma^2} \right)^{\frac{\zeta^2}{\sigma^2}} }{\Gamma(\frac{\zeta^2}{\sigma^2})}
\label{eq:lik_y}
\end{align}

and, hence, the log-likelihood is given by
\begin{align}
\ell(\zeta,\sigma; \mathbf{y}) & = \sum_{i=1}^n \left( \left(\frac{\zeta^2}{\sigma^2}-1\right)log(y_i)-\frac{\zeta}{\sigma^2} y_i + \frac{\zeta^2}{\sigma^2} log\left(\frac{\zeta}{\sigma^2}\right) - log \left( \Gamma\left(\frac{\zeta^2}{\sigma^2}\right) \right) \right) \nonumber \\
& = n \times \frac{\zeta^2}{\sigma^2} log\left(\frac{\zeta}{\sigma^2}\right)-n \times log\left(\Gamma\left(\frac{\zeta^2}{\sigma^2}\right)\right)+\sum_{i=1}^{n} \left( \left(\frac{\zeta^2}{\sigma^2}-1\right)log(y_i)-\frac{\zeta}{\sigma^2} y_i \right) 
\end{align}
where $n$ represents the dimension of all data ($n=\sum_{t=t_1}^{t_T} (n^S(t)+n^C(t))$)

\subsection{Likelihood for $\mathbf{Z} \vert \mathbf{X},\mathbf{X}^{S},\mathbf{V},\mathbf{W}$}

\begin{align}
\mathcal{L}(\pi; \mathbf{z}) = \prod_{i=1}^{n} \pi^{z_i}(1-\pi)^{1-z_i}
= \pi^{\sum_{i=1}^{n}z_i}(1-\pi)^{n-\sum_{i=1}^{n}z_i}.
\label{eq:lik_z}
\end{align}

The corresponding log-likelihood is given by
\begin{align}
\ell(\pi; \mathbf{z})  = log(\pi)\sum_{i=1}^n z_i + log(1-\pi) \left(  n-\sum_{i=1}^n z_i \right).
\end{align}

\subsection{Likelihood for $\mathbf{X} \vert \mathbf{U}, \mathbf{V} $}

Following \cite{Diggle2013}, the likelihood for IPP comes from
\begin{align}
\mathcal{L}(\lambda({.,t}); \mathbf{x}) = & \prod_{i=1}^{n^C(t)} \frac{e^{-\omega}\omega^{n^C(t)}}{n^C(t)!}\times \frac{\lambda({\mathbf{x}_i^C,t)}}{\omega}.
\label{eq:lik_x}
\end{align}

Thus, the log-likelihood is expressed as
\begin{align}
\ell(\lambda({.,t}); \mathbf{x})  = & ~log \left( \frac{e^{-\omega}\omega^{n^C(t)}}{n^C(t)!} \right)+ \sum_{i=1}^{n^C(t)}
log \left( \frac{\lambda({\mathbf{x}_i,t})}{\omega} \right) \nonumber\\ = & -\omega + n^C(t) \times log(\omega)-log(n^C(t)!)+\sum_{i=1}^{n^C(t)} \left( log(\lambda({\mathbf{x}_i,t}))-log(\omega) \right)\nonumber \\ \simeq  & \sum_{i=1}^{n^C(t)} log(\lambda({\mathbf{x}_i,t}))-\omega = \sum_{i=1}^{n^C(t)} log(\lambda({\mathbf{x}_i,t}))-\int_{\mathcal{A}} \lambda({\mathbf{s},t})\partial \mathbf{s}.
\end{align}

\subsection{Likelihood for $\mathbf{W}$}

\begin{align}
\mathcal{L}(\sigma_W^2,\phi_W,\delta) = & \frac{1}{(2\pi)^{N/2} \vert (1-\delta^2)^{-1}\Sigma_W \vert^{1/2}}exp \left\lbrace -\frac{1}{2}\mathbf{w}'({.,t_1}) \left( (1-\delta^2)^{-1}\Sigma_W \right)^{-1} \mathbf{w}({.,t_1}) \right\rbrace  \nonumber \\
 &+ \sum_{t=t_2}^{t_T} \frac{1}{(2\pi)^{N/2} \vert \Sigma_W \vert^{1/2}}exp \left\lbrace -\frac{1}{2}(\mathbf{w}'({.,t})-\delta\mathbf{w}'({.,t-1})) \Sigma_W^{-1} (\mathbf{w}({.,t})-\delta\mathbf{w}({.,t-1})) \right\rbrace
\label{eq:lik_w}
\end{align}

and, hence, the log-likelihood is given by

\begin{align}
\ell(\sigma_W^2,\phi_W,\delta) = & -\frac{N}{2}log(2\pi)-\frac{N}{2}log(\vert (1-\delta^2)^{-1}\Sigma_W \vert) -\frac{1}{2}\mathbf{w}'({.,1}) \left( (1-\delta^2)^{-1}\Sigma_W \right)^{-1} \mathbf{w}({.,1}) \nonumber \\
& +\sum_{t=t_2}^{t_T} (  -\frac{N}{2}log(2\pi) - \frac{1}{2}log(\vert \Sigma_W \vert ) - \frac{1}{2}(\mathbf{w}'(.,t)-\delta\mathbf{w}'({.,t-1})) \Sigma_W^{-1} (\mathbf{w}({.,t})-\delta\mathbf{w}(.,t-1))  ) \nonumber \\
= & -\frac{NT}{2}log(2\pi) +\frac{N}{2}log(1-\delta^2)-\frac{T}{2}log(\vert \Sigma_W \vert)\nonumber \\
& - \frac{1}{2}\mathbf{w}'(.,1) \left( (1-\delta^2)^{-1}\Sigma_W \right)^{-1} \mathbf{w}(.,1)\nonumber \\
&- \sum_{t=t_2}^{t_T} \left(\frac{1}{2}(\mathbf{w}(.,t)-\delta\mathbf{w}'(.,t-1)) \Sigma_W^{-1} (\mathbf{w}(.,t)-\delta\mathbf{w}(.,t-1))\right).
\end{align}

\subsection{Likelihoods for $\mathbf{U}$ and $\mathbf{V}$}
\begin{align}
\mathcal{L}(\sigma_l, \phi_l) = & \frac{1}{(\sqrt{2\pi})^N \vert \Sigma_U \vert ^{1/2}}exp \lbrace -\frac{1}{2}\mathbf{l}'(.) \Sigma_U^{-1} \mathbf{l}(.) \rbrace,
\label{eq:lik_u_v}
\end{align}

and, hence, the log-likelihood is given by
\begin{align}
\ell(\sigma_l, \phi_l)  = & -\frac{N}{2}log(\pi) -\frac{log(\vert \Sigma_l \vert)}{2} - \frac{1}{2}\mathbf{l}'(.) \Sigma_l^{-1} \mathbf{l}(.),
\end{align}
where $N$ denotes the dimension of the prediction grid (or mesh), $l=\{\mathbf{U},\mathbf{V}\}$ and $\mathbf{l}=\{\mathbf{u},\mathbf{v}\}$.

\newpage

\section{Scenarios of sampling}
$~$
\begin{figure}[H]
    \centering
    \includegraphics[width=1\textwidth]{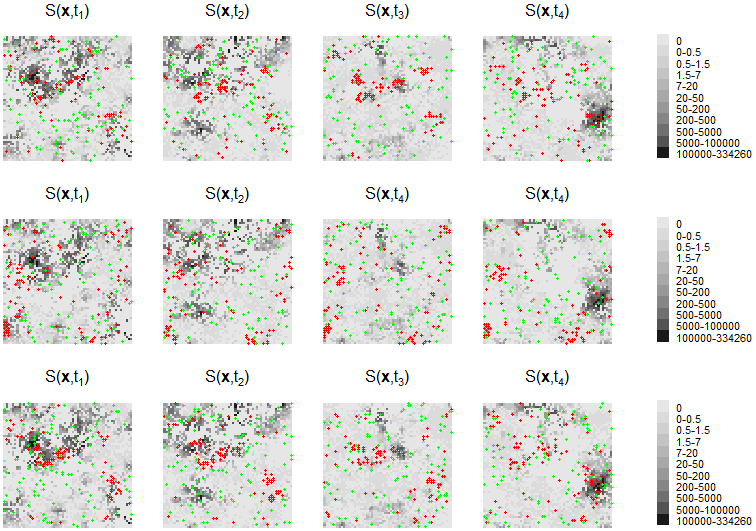}
    \caption{Examples of simulated both FID (fishery-independent data) and FDD (fishery-dependent data) locations across sampling scenarios: \textit{Scenario 1} ($\beta'=0$ and $\beta(t) \sim N(2,0.25)$), \textit{Scenario 2} ($\beta'(t) \sim N(2,0.25)$ and $\beta(t)=0$), and \textit{Scenario 3} ($\beta'(t) \sim N(1,0.25)$ and $\beta(t) \sim N(2,0.25)$) with $t=\{t_1,\cdots,t_4\}$. \textcolor{green}{Green} points represent sample locations for simulated FID, and \textcolor{red}{red} points identify the sample locations for simulated FDD at time $t$. The depicted latent field is $\mathbf{S}$, and each data source has a dimension of 100.}
    \label{fig:scenarios}
\end{figure}

\newpage
\section{Supplementary results of the simulation study}
\subsection{Evaluation of the estimation of intercept and regression coefficient parameters}
\begin{figure}[h!]
    \centering
    \begin{subfigure}{0.49\linewidth}
        \includegraphics[width=\linewidth]{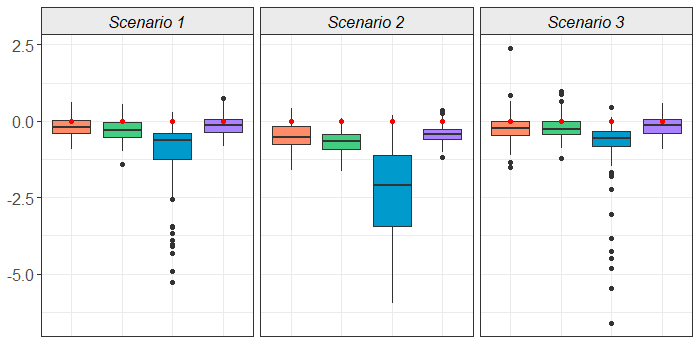}\\
        \includegraphics[width=\linewidth]{combinations_legend.png}
        \caption{$\alpha'$}
        \label{fig:alpha_z}
    \end{subfigure}
    \begin{subfigure}{0.49\linewidth}
        \includegraphics[width=\linewidth]{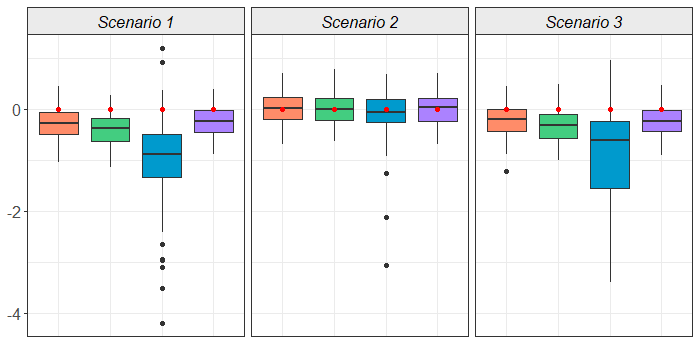}\\
        \includegraphics[width=\linewidth]{combinations_legend.png}
        \caption{$\alpha$}
        \label{fig:alpha_y}
    \end{subfigure}
    \caption{Estimates of intercept parameters ($\alpha'=0$ and $\alpha=0$) across sampling scenarios and combinations of samples’ dimensions $Comb(n^{I}(t),n^{D}(t))$.  Scenarios are defined as: \textit{Scenario 1} with $\beta'(t) = 0$ and $\beta(t) \sim N(2,0.25)$, \textit{Scenario 2} with $\beta'(t) \sim N(2,0.25)$ and $\beta(t) = 0$, and \textit{Scenario 3} with $\beta'(t) \sim N(1,0.25)$ and $\beta(t) \sim N(2,0.25)$ for $t=\{t_1,\cdots,t_4\}$. \textcolor{red}{Red} points represent the true values of the corresponding parameter for all 100 replicas.}
    \label{fig:alpha_by_comb_dim}
\end{figure}
\begin{figure}[h!]
    \centering
    \includegraphics[width=0.75\linewidth]{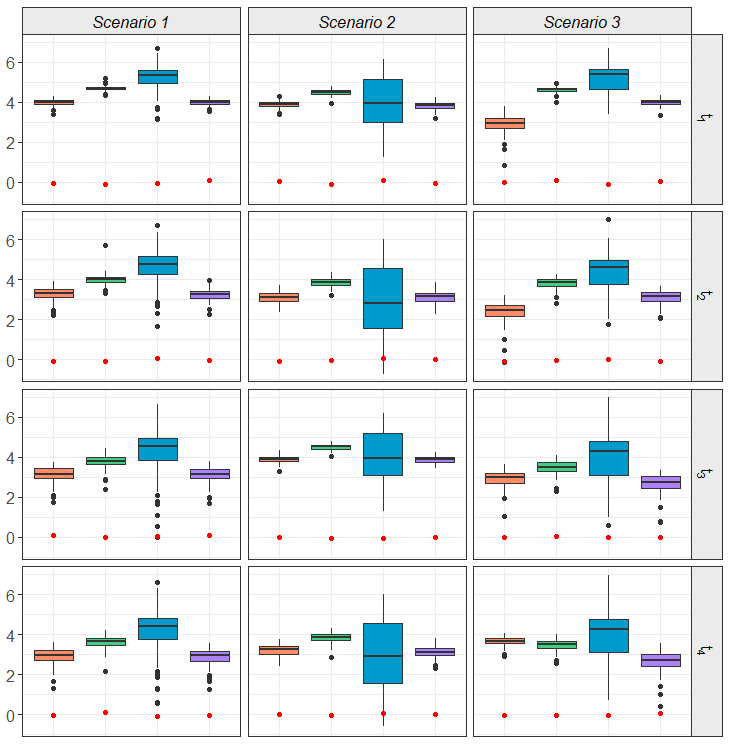}\\
    \includegraphics[width=0.75\linewidth]{combinations_legend.png}
    \caption{Estimates of $\alpha_t''=0$ across sampling scenarios (\textit{Scenario 1}: $\beta'_t = 0$ and $\beta_t \sim N(2,0.25)$; \textit{Scenario 2}: $\beta'_t \sim N(2,0.25)$ and $\beta_t = 0$; \textit{Scenario 3}: $\beta'_t \sim N(1,0.25)$ and $\beta_t \sim N(2,0.25)$), combinations of samples’ dimensions $Comb(n_t^{I},n_t^{D})$, and time units $t=\{t_1,\cdots,t_4\}$. The \textcolor{red}{red} points represent the true values of the corresponding parameter for all 100 replicas.}
    \label{fig:alpha_intens}
\end{figure}

$~$
\newpage

$~$
\newpage

\subsection{Evaluation of the predictive performance}
\begin{figure}[!h]
    \centering
    \begin{subfigure}{0.49\linewidth}
    \centering
        \includegraphics[width=\linewidth]{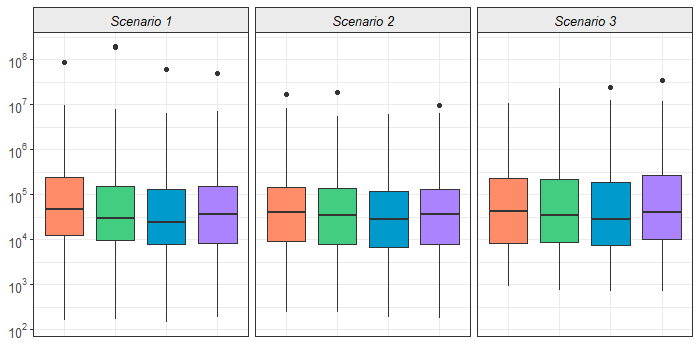}\\
        \includegraphics[width=\linewidth]{combinations_legend.png}
        \caption{RMSE (Root Mean Square Error)}
        \label{fig:rmse}
    \end{subfigure}
    \begin{subfigure}{0.49\linewidth}
    \centering
        \includegraphics[width=\linewidth]{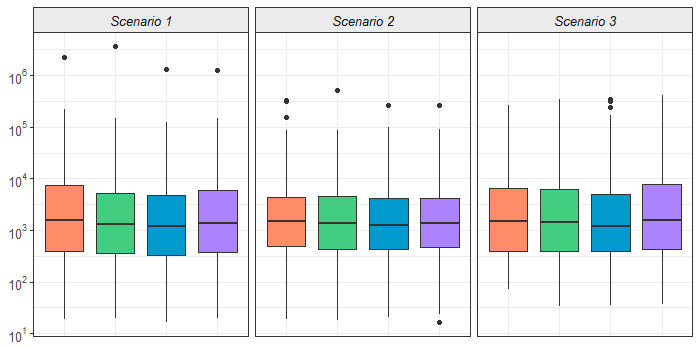}\\
        \includegraphics[width=\linewidth]{combinations_legend.png}
        \caption{MAE (Mean Absolute Error)}
        \label{fig:mae}
    \end{subfigure}
    \caption{Evaluation of predictive performance. Performance metrics (RMSE and MAE) across sampling scenarios and combinations of samples' dimensions $Comb(n^{I}(t),n^{D}(t))$. Scenarios are defined as: \textit{Scenario 1} with $\beta'(t) = 0$ and $\beta(t) \sim N(2,0.25)$, \textit{Scenario 2} with $\beta'(t) \sim N(2,0.25)$ and $\beta(t) = 0$, and \textit{Scenario 3} with $\beta'(t) \sim N(1,0.25)$ and $\beta(t) \sim N(2,0.25)$ for $t=\{t_1,\cdots,t_4\}$. The results are scaled using $log_{10}$ transformation.}
    \label{fig:model_perf}
\end{figure}
$~$

\newpage
\subsection{Evaluation of the contribution of each data source}
$~$
\begin{figure}[!h]
 \centering
    \begin{subfigure}{0.49\linewidth}
    \centering
        \includegraphics[width=\linewidth]{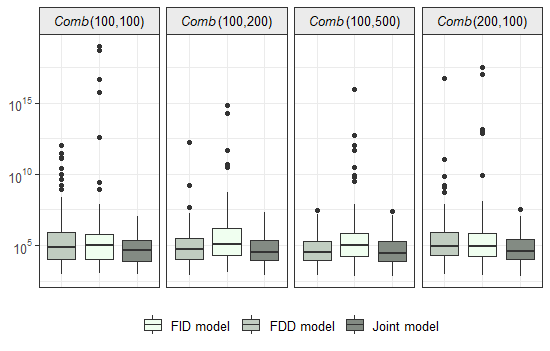}
        \caption{RMSE (Root Mean Square Error)}
        \label{fig:rmse_3models}
    \end{subfigure}
    \begin{subfigure}{0.49\linewidth}
    \centering
        \includegraphics[width=\linewidth]{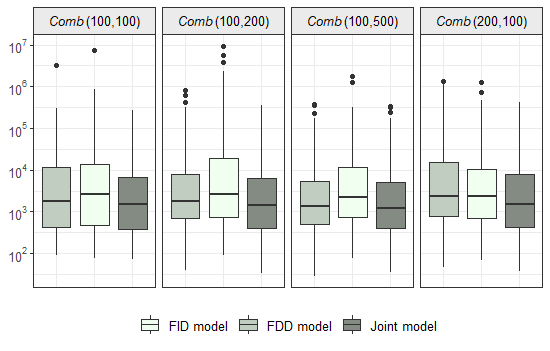}
        \caption{MAE (Mean Absolute Error)}
        \label{fig:mae_3models}
    \end{subfigure}
    \caption{Evaluation of the contribution of each data source. Performance metrics (RMSE and MAE) across  model configurations of sample dimension $Comb(n^{I}(t),n^{D}(t))$ under \textit{Scenario 3} ($\beta'(t)\sim N(1,0.25)$ and $\beta(t)\sim N(2,0.25)$) for $t=\{t_1,\cdots,t_4\}$. The results are scaled using $log_{10}$ transformation.}
    \label{fig:model_perf_3models}
\end{figure}

\newpage
\section{Model performance results with varying the catchability effects}
\begin{table}[H]
\centering
\caption{Model performance based on AIC (Akaike Information Criteria) with varying definitions of catchability effect $k(v)$.
AIC values for models with different definitions of catchability effects, assessing the impact of incorporating vessel-specific parameters versus individual vessel features. Lower AIC values indicate better model fit, guiding the final model choice to use a simplified catchability parameter for each vessel without specific vessel features. $Power(v)$ and $Length(v)$ denote the engine power and the length of vessel $v$. $\gamma_c(v)$ represents the random effect for each vessel. $f$ identifies the smoother therm ($bs$: B-spline basis; $cr$: cubic regression spline basis) with $knots$ knots.}
\small
\begin{tabular}{lc}
\hline
Expression for catchability effect $k(v)$                                            & AIC    \\ \hline
$\alpha_c$                                     & 161.42 \\
$exp \{\alpha_c + f(Power(v),therm=bs,knots=7) + \gamma_c(v)\}$  & 176.56 \\
$exp \{\alpha_c + f(Power(v),therm=bs,knots=11) + \gamma_c(v)\}$ & 175.36 \\
$exp \{\alpha_c + f(Power(v),therm=cr,knots=5) + \gamma_c(v)\}$  & 174.50 \\
$exp \{\alpha_c + f(Length(v),therm=cr,knots=7) + \gamma_c(v)\}$ & 174.13 \\
$exp \{\alpha_c + \gamma_c(v)\}$                    & 168.83 \\ \hline
\end{tabular}
\end{table}
\normalsize

\end{document}